\documentclass[9pt,%
 twocolumn,
,amssymb,
floatfix,
]{revtex4-2}
\usepackage{lipsum}
\usepackage{color}
\usepackage{afterpage}
\usepackage{amsmath}
\usepackage{graphicx}
\usepackage{float}

\usepackage{dcolumn}
\usepackage{bm}
\usepackage{amsmath}
\begin{document}

\title{Tunable Three-Dimensional Architecture of Nematic Disclination Lines}

\author{Alvin Modin$^{1}$}
\thanks{Authors contributed equally.}

\author{Biswarup Ash$^{2}$} 
\thanks{Authors contributed equally.}

\author{Kelsey Ishimoto$^{1}$}
\author{Robert L. Leheny$^{1}$}
\author{Francesca Serra$^{3}$}
\author{Hillel Aharoni$^{2}$}
\affiliation{$^1$Department of Physics and Astronomy, Johns Hopkins University, Baltimore, MD 21218, USA}
\affiliation{$^2$Department of Physics of Complex Systems, Weizmann Institute of Science, Rehovot 76100, Israel}
\affiliation{$^3$Department of Physics, Chemistry, and Pharmacy, University of Southern Denmark, Odense, DK-5230 Denmark}

\begin{abstract}

Disclinations lines play a key role in many physical processes, from the fracture of materials to the formation of the early universe. Achieving versatile control over disclinations is key to developing novel electro-optical devices, programmable origami, directed colloidal assembly, and controlling active matter. Here, we introduce a theoretical framework to tailor three-dimensional disclination architecture in nematic liquid crystals experimentally. We produce quantitative predictions for the connectivity and shape of disclination lines found in nematics confined between two thinly spaced glass substrates with strong planar anchoring. By drawing an analogy between nematic liquid crystals and magnetostatics, we find that: i) disclination lines connect defects with the same topological charge on opposite surfaces, and ii) disclination lines are attracted to regions of the highest twist. Using polarized light to pattern the in-plane alignment of liquid crystal molecules, we test these predictions experimentally and identify critical parameters that tune the disclination lines' curvature. We verify our predictions with computer simulations and find non-dimensional parameters enabling us to match experiments and simulations at different length scales. Our work provides a powerful method to understand and practically control defect lines in nematic liquid crystals.
\end{abstract}
\maketitle

Topological singularities link physically distinct phenomena -- they mediate phase transitions \cite{kosterlitz_ordering_1973}, act as organizational centers in biological systems \cite{maroudas-sacks_topological_2021}, and steer the trajectory of light \cite{brasselet_tunable_2012, meng_topological_2023}. Various topological defect configurations are present in nematic liquid crystals (LCs), fluid-like materials with long-range orientational molecular order. Disclination lines arise when nematic LCs are frustrated by incompatible boundary conditions. These one-dimensional singularities can be facilely formed and visualized, making nematic LCs an ideal test bed for studying defect structures and interactions. Manipulating disclination lines in LCs also has practical applications in directed self-assembly, \cite{blanc_ordering_2013, luo_tunable_2018}, tunable photonics \cite{coles_liquid_2005}, and re-configurable microfluidic devices \cite{sengupta_topological_2013, sandford_oneill_electrically-tunable_2020}.  To effectively utilize the potential of disclinations for these applications, it is essential to develop a set of fundamental
rules that govern their formation and connectivity.

Recent advances in spatial patterning of liquid crystal alignment have enabled greater control over the structure of disclination lines. For example, imprinted nano-ridges on glass substrates have been used to precisely shape defect lines, revealing insights into their energy, structure, and multi-stability \cite{chuck,harkai_electric_2020}. Using light to impose LC alignment at photosensitive substrates is an equally powerful tool. Photo-alignment has enabled the design of free-standing disclination loops\cite{wang_artificial_2017, ouchi_topologically_2019,sunami_shape_2018} and periodic disclination arrays with different morphologies and properties \cite{guo_photopatterned_2021, yoshida_three-dimensional_2015,nys_nematic_2022,nys_periodic_2018,berteloot_ring-shaped_2020,jiang_active_2022}. 

\begin{figure}[t]
\centering
\includegraphics[width=0.92\columnwidth]{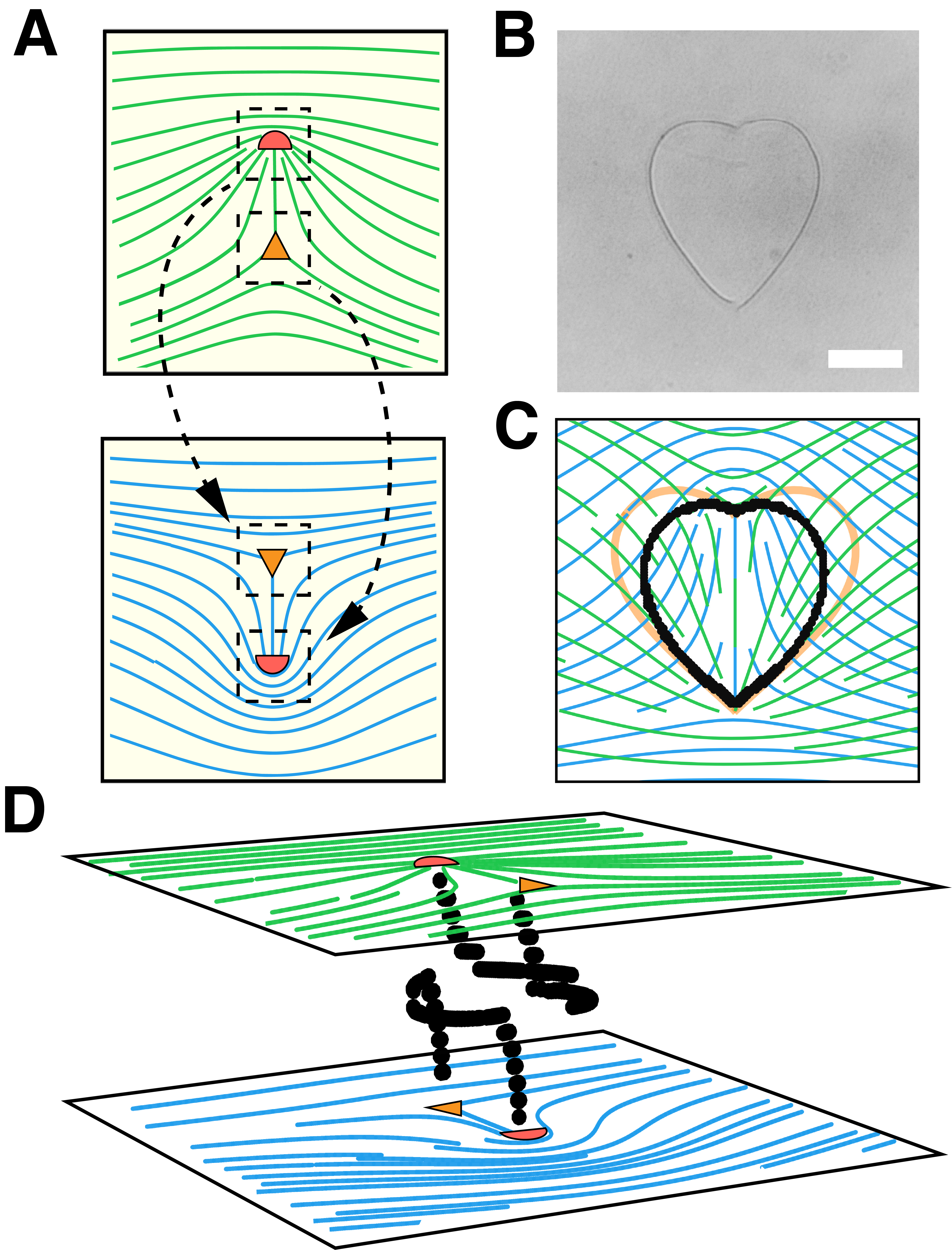}
\caption{ \textbf{Designing three-dimensional disclination line architecture.} ({\textit A}) Schematics of the two-dimensional surface patterns imprinted onto substrates coated with a photo-sensitive layer (Brilliant Yellow). Topological defects with charge $+1/2$ (red semi-circle) and  $-1/2$ (orange triangle) act as nucleation sites for disclination lines. The substrates are aligned so that defects of opposing strength are in registry with each other. ({\textit B}) Disclination lines viewed under bright-field microscopy form a heart-shaped structure (scale bar: 25 $\mu \rm{m}$). This configuration corresponds to a stable equilibrium state.  Top view ({\textit C}) and side view ({\textit D}) of the disclination structure obtained from simulation using the patterns in ({\textit A}) as surface boundary conditions. The observed two-dimensional projection (black points) is a balance between the attraction of defect lines to locations where the top and bottom surface patterns are perpendicular (orange lines in ({\textit C})) and their line tension. ({\textit D}) A side view of the defect configuration obtained in the simulation reveals that the disclination lines primarily run through the mid-plane of the cell and connect surface defects of opposite charge on the same substrate.}

\label{fig:defectArchitecture}
\end{figure}

In this work, we introduce a general framework for creating arbitrarily shaped three-dimensional (3D) disclination line architecture in nematic liquid crystals. As an example, we show a structure where the projection of disclination lines on a two-dimensional (2D) plane forms the shape of a heart (Fig.~\ref{fig:defectArchitecture}). In the experiment, light-sensitive layers on parallel glass substrates align the nematic at the surfaces in patterns decorated with pairs of 2D surface defect nucleation sites (Fig.\ref{fig:defectArchitecture}\textit{A}). The 2D defects are characterized by winding numbers - the degree of rotation of the nematic director around the defect divided by 2$\pi$ - of $+1/2$ and $-1/2$. Aligning opposite-charged defects on opposing substrates, we observe that the confined LC forms a pair of disclination lines that primarily run through the mid-plane of the cell to connect surface defects on the same substrate (Fig. \ref{fig:defectArchitecture}\textit{B}). This configuration is a stable, equilibrium state, confirmed by numerical simulations (Fig.\ref{fig:defectArchitecture}\textit{C-D}).

To understand the paths that the disclination lines take, we draw an analogy between the elastic distortion of a nematic and the magnetostatic field of current-carrying wires. Using this analogy, we  experimentally and numerically verify two key rules: (i) disclination lines either connect surface defects on opposing substrates with the same winding number or surface defects on the same substrate having opposite winding numbers; (ii) the lines' paths depend on the interplay between forces driving them to regions of maximum twist set by the confining pattern and the disclination line tension. We utilize these two design principles to create the heart-shaped disclinations shown in Fig.~\ref{fig:defectArchitecture}. Our proposed framework enables the design of tunable 3D liquid crystal-based disclination networks for applications in re-configurable optics, photonic devices, and responsive matter.

The following sections explain the magnetostatic analogy and its outcomes. Subsequently, we test specific predictions with experiments and simulations. Finally, we revisit the structure in Fig.~\ref{fig:defectArchitecture} to explain how we designed the heart-shaped lines and how disclination shapes can be tailored by varying temperature.
\vspace{-4mm}
\section*{Results and Discussion}
\subsection*{Magnetostatics model}
Distortions of the nematic director field are described by the Frank-Oseen elastic free energy \cite{deGennesBook},

\begin{equation}
    \begin{split}
    	F_{\textrm{el}}=\int \left[\frac{K_1}{2}\left(\nabla\cdot\mathbf{\hat{n}}\right)^2+\frac{K_2}{2}\left(\mathbf{\hat{n}}\cdot\nabla\times\mathbf{\hat{n}}\right)^2\right.+\\ \left.\frac{K_3}{2}\left(\mathbf{\hat{n}}\times\nabla\times\mathbf{\hat{n}}\right)^2 \right]dV,
    \label{FrankOseen}
    \end{split}
\end{equation}
with $\mathbf{\hat{n}}$ the nematic director and $K_{1,2,3}$ the splay, twist, and bend elastic constants, respectively. We consider a nematic placed between two parallel plates with (sufficiently strong) planar anchoring on them, separated by a spacing $t$ much smaller than their lateral dimensions. Under these conditions, we make the following key assumption: in equilibrium, the nematic director is planar everywhere within the cell, not only at the boundaries. This is analogous to the Kirchhoff-Love assumptions in plate elasticity theory \cite{Love1888}. The nematic director field then takes the form $\mathbf{\hat{n}}=\left(\cos\theta,\sin\theta,0\right)$, where $\theta(x,y,z)$ is the director's azimuthal angle in the $xy$-plane. In addition, we use the two-constant approximation, with $K_1=K_3 \equiv K$, which is valid near the nematic-isotropic transition for low molecular weight thermotropic nematic LCs, particularly  4'-octyl-4-biphenylcarbonitrile (8CB) \cite{elasticConsts} used in our experiments. Eq. \ref{FrankOseen} then assumes the simple form,
\begin{equation}
\label{FreeEnergy}
	F_{el}=\int\left[\frac{K}{2}\left[(\partial_{x}\theta)^2+(\partial_{y}\theta)^2\right]+\frac{K_2}{2}(\partial_{z}\theta)^2\right]dV.
\end{equation}

Further simplification is obtained by rescaling the $z$-axis using $\tilde{z}=z\sqrt{K/K_2}$ (defined on a domain of thickness $\tilde{t}=t\sqrt{K/K_2}$), and redefining $\nabla\equiv(\partial_x,\partial_y,\partial_{\tilde{z}})$, so that
\begin{equation}
	F_{el}=\frac{K}{2}\int\left|\nabla\theta\right|^2 d\tilde{V}.
\label{FreeEnergySimplest}
\end{equation}
The functional in Eq.~\ref{FreeEnergySimplest} implies that, in equilibrium, $\theta(x,y,\tilde{z})$ is a harmonic function. However, this property breaks down along disclination lines; at the defect core, the nematic order vanishes, and $\theta$ is not defined. Around the defect line, Eq.~\ref{FreeEnergySimplest} admits a nontrivial quantized integral,

\begin{equation}\label{disclination}
	\oint d\mathbf{\ell}\cdot\nabla\theta=2\pi q,\qquad q\in\mathbb{Z}/2.
\end{equation}

Together Eq.~\ref{FreeEnergySimplest} and \ref{disclination} establish an exact mathematical analogy of the nematic cell to magnetostatics, as previously identified by de Gennes \cite{deGennesBook}. In the analogy, the planar director's azimuthal angle $\theta$ plays the role of a magnetic scalar potential, whose gradient is the magnetic field. Disclination lines are current-carrying wires. Their existence renders $\theta$ ambiguous; however, the half-integer quantization of the current exactly corresponds to the nematic $\theta\cong\theta+\pi\mathbb{Z}$ congruence.

The disclination wires are flexible and stretchy. Each wire is associated with a line tension $\gamma$, the outcome of melting of the nematic order at the defect core to alleviate the diverging elastic energy. Approximately, $\gamma$ is proportional to $K q^2$; however, there are logarithmic corrections that depend on the line and cell geometry \cite{deGennesBook,Schopohl1987}. These corrections become significant near the nematic-isotropic phase transition as the defect core size diverges. For simplicity, we ignore these corrections and treat $\gamma$ as a constant. Similarly to other material parameters, namely $K, K_2$, $\gamma$ may depend on temperature in a non-trivial way.

\subsubsection*{Forces on wires}
To study the shape of disclination wires, we calculate the effective forces acting on them (see \textit{SI Appendix} for full derivations). There are three forces (per unit length) acting on the wires:
\begin{enumerate}
		\item The strong anchoring on the two surfaces acts as magnetic mirrors. Disclination wires are repelled by these mirrors (alternatively, by the mirror image wires) and pushed toward the mid-plane between the two boundary surfaces by a force
		\begin{equation}
			\mathbf{f}_M=-\frac{\pi^2 K q^2}{\tilde{t}}\tan\left(\frac{\pi \tilde{z}}{\tilde{t}}\right)\mathbf{\hat{z}}.
        \label{eq:fmirrors}
		\end{equation}
		\item The anchoring planar angles, $\theta_{t,b}(x,y)$ on the top/bottom surfaces, respectively, are analogous to an external magnetic field that exerts a Lorentz-like force on disclination wires:
		\begin{equation}
			\mathbf{f}_B=\frac{2\pi K q}{\tilde{t}} \left(\theta_{t}-\theta_{b}-q\pi\right)\mathbf{\hat{T}}\times\mathbf{\hat{z}},
        \label{eq:fexternal} 
		\end{equation}
		where $\mathbf{\hat{T}}$ is the unit tangent to the defect line. This force pulls defect lines horizontally towards regions where the top and bottom are at a $\Delta \theta \equiv \theta_t - \theta_b = q\pi$ angle difference from each other.
		\item The line tension of the wires exerts a force
		\begin{equation}
            \mathbf{f}_\gamma=\gamma\kappa\mathbf{\hat{N}},
        \label{eq:ftension}
		\end{equation}
		where $\kappa$ is the curvature of the wire and $\mathbf{\hat{N}}$ its normal in the Frenet-Serret frame.
\end{enumerate}
The equilibrium shape of a disclination line is obtained by the balance of $\mathbf{f}_M$, $\mathbf{f}_B$, and $\mathbf{f}_{\gamma}$.  

\begin{figure*}[t!]
\centering
\includegraphics[width=1\textwidth]{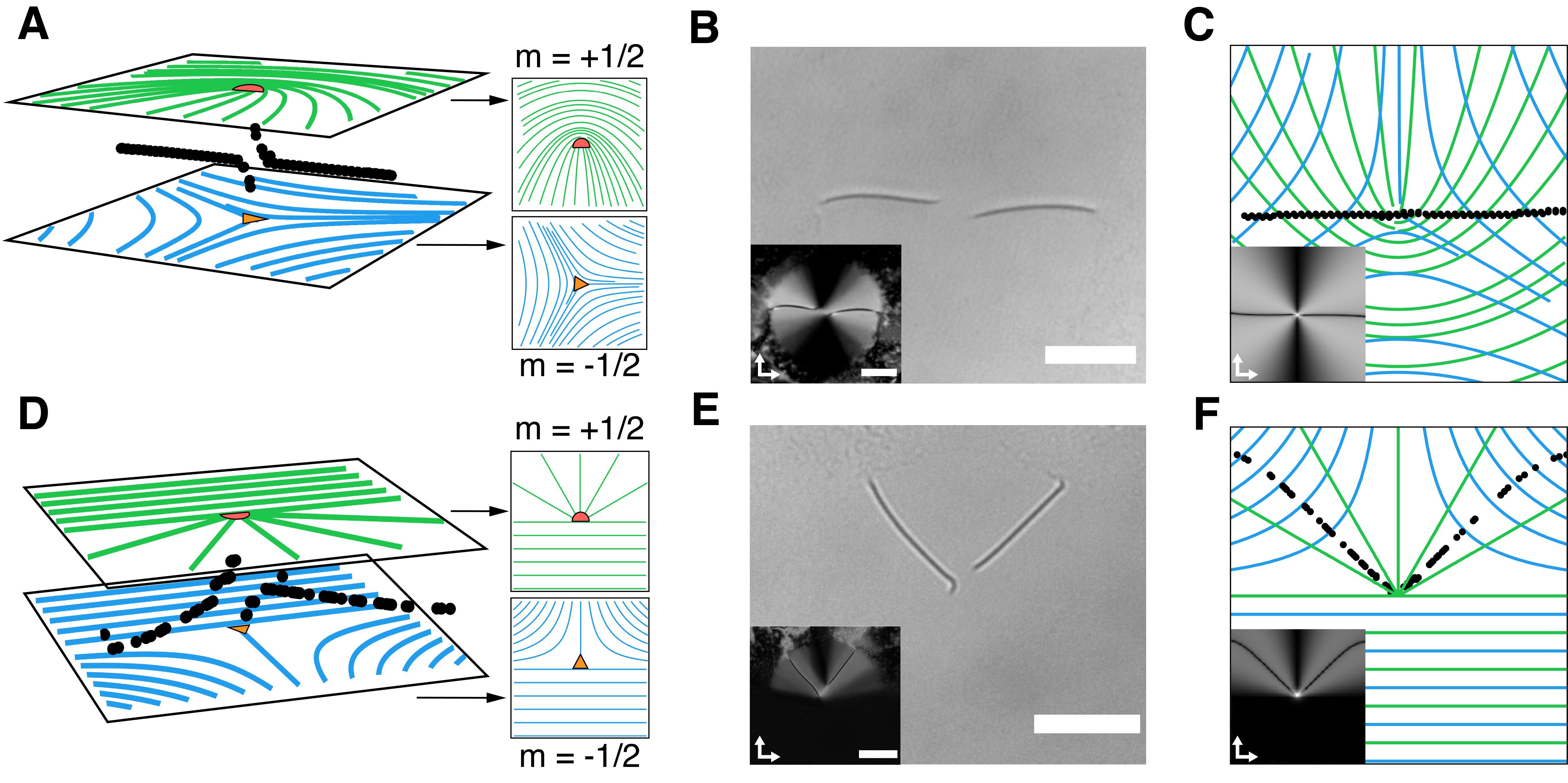}
\caption{\textbf{Surface defect connectivity.} ({\textit A}) When two canonical surface defects with $\pm 1/2$ charge are imposed as boundary conditions, the planarity of the director field forbids the connection of the top and bottom surface defects with a single defect line. The equilibrium state obtained from numerical minimization of Landau de Gennes energy corresponds to two disclination lines (black points) nucleating at the surface defect cores and extending towards the boundaries.({\textit B})  Bright-field microscopy images where two confining surfaces are photo-aligned with isolated $\pm 1/2$ surface defects corresponding to the pattern used in ({\textit A}) . Inset: polarized light microscopy image.({\textit C}) Two-dimensional projection of the numerical results in ({\textit A}). Inset: The expected polarized optical microscopy (POM) texture reconstructed from the director field using Jones calculus. ({\textit D}) Preserving the topological charge of the surface defects while altering their geometric structure changes the regions where $\Delta \theta = \pi/2$. The equilibrium state for these boundary conditions corresponds to two disclination lines now at a relative angle of $\pi/2$. ({\textit E}) Bright-field microscopy images where two confining surfaces are photo-aligned with isolated $\pm 1/2$ surface defects corresponding to the pattern used in ({\textit D}). ({\textit F}) Two-dimensional projections of the numerical results in ({\textit D}). Inset: The expected POM texture reconstructed from the director field using Jones calculus. Scale bars: 25$\mu$m.} 
\label{fig:defectConnectivity}
\end{figure*}


This force balance can characterize the geometry of defect lines connecting to surface defects. A disclination line emerges perpendicularly from a surface topological defect due to the magnetic repulsion described by Eq.~\ref{eq:fmirrors}. For a surface defect $|q|>\frac{1}{2}$ we expect a split into $2|q|$ ``atomic'' lines of magnitude $\frac{1}{2}$ arising from the mutual repulsion between them (as was observed in \cite{yoshida_three-dimensional_2015, chuck}). By the balance of forces, $\mathbf{f}_M$ and $\mathbf{f}_\gamma$ lines emerging from defects turn horizontally into the mid-plane over a typical length scale $\sim\frac{\gamma}{K}\tilde{t}\equiv\frac{\gamma}{\sqrt{K K_2}}t$. Thus, defect lines whose lateral span is much larger than this scale traverse within the mid-plane for the more significant part of their trajectories.

\subsubsection*{Connectivity of surface defects}

We study the topological rules of connecting -- with disclination wires -- surface defects patterned on two confining surfaces. Each surface defect is characterized by a winding number $q$, defined by a closed loop $\Gamma$ around the defect core as $\frac{1}{2\pi}\oint_\Gamma \nabla\theta\cdot d\ell$. By current conservation, a disclination line can connect two surface defects of the \textit{same $q$ on opposite surfaces}; or two surface defects of \textit{opposite $q$ on the same surface}. Alternatively, disclinations can escape to the sides of the system or form a closed loop. Planarity of the director field forbids connection between a top-surface $+1/2$ defect with a bottom-surface $-1/2$, even though this would be topologically allowed in a 3D nematic liquid crystal \cite{mermin_topological_1979}.
\\

\subsection*{Experimental Tests}
\subsubsection*{Test of Design Principles}
To verify these connectivity principles experimentally, we create a LC cell where the bottom and top surfaces contain a single, isolated $q = -1/2$ or $+1/2$ defect, respectively (Fig.~\ref{fig:defectConnectivity} \textit{A-C}), utilizing the custom built photo-alignment system described in \textit{Materials and Methods} and \textit{SI Appendix, Fig. S1} \cite{chigrinov_photoalignment_2008,folwill_practical_2021, priimagi}. We shine linearly polarized light on glass coated with a light-sensitive alignment layer (Brilliant Yellow). The alignment layer molecules give planar alignment to the LC, with a direction that is perpendicular to the polarization of the incident light. By spatially patterning the light polarization, we imprint half-integer defect nucleation sites onto confining glass substrates. The defects on each surface are photo-patterned within a circular patch of diameter $d \approx 75 \mu$m.  Under crossed-polarizers, the dichroic properties of the Brilliant Yellow dye enable us to view the patterned regions on the confining glass substrates before filling them with LC. We align the circular patches on each substrate to overlap, ensuring that defect cores of opposing topological charges are in registry. Once substrates are secured with epoxy resin, we carefully measure the cell thickness and inject pre-heated 8CB LC into the cell, allowing it to slowly cool until it reaches the nematic phase at $\sim 36^\circ \rm{C}$.

The resulting defect structure follows the connectivity rules: rather than a single disclination line connecting the surface defects as might be expected \cite{mermin_topological_1979}, two disclination lines emerge from the defect cores and escape to the sides along the mid-plane, as can be seen from a side view of the numerical simulation in Fig.~\ref{fig:defectConnectivity}\textit{A} and from the top view in experiments (Fig.~\ref{fig:defectConnectivity}\textit{B}) and simulations (Fig.~\ref{fig:defectConnectivity}\textit{C}). Indeed, this connectivity rule gives rise to the two extended lines that make the heart shape in Fig.~\ref{fig:defectArchitecture} rather than two defect lines connecting surface defects directly facing each other. For verification, we run the same experiment and simulation with $+1/2$ surface defects patterned onto confining substrates (see \textit{SI Appendix, Fig. 2}). A vertical disclination line connects the top and the bottom surface defects, as is permitted in this case by current conservation. 

The effect of varying the patterned boundary conditions can be seen in Fig.~\ref{fig:defectConnectivity}(\textit{D-F}). Here, we preserve the topological charge of each surface pattern but introduce a homogeneously aligned region that alters the geometric structure of the $\pm 1/2$ defects. The new surface pattern modifies the areas where $\Delta \theta \left(x,y\right)$ are orthogonal. The new regions where $\mathbf{f}_{B} = 0$ lead to a reduced angular separation of the two defect lines from $\pi$ (Fig.~\ref{fig:defectConnectivity}\textit{B-C}) to $\pi/2$ (Fig.~\ref{fig:defectConnectivity}\textit{E-F}). For both designs, the disclination wires do not bend and $\kappa \approx 0$, implying that $\mathbf{f}_\gamma$ has little to no effect on the positioning of the lines.

\subsubsection*{Tuning the curvature of disclination architecture}
When the imposed surface patterns result in curved disclination lines, the line tension $\mathbf{f}_\gamma$ becomes important. $\mathbf{f}_\gamma$ opposes $\mathbf{f}_B$, acting to minimize the wire's curvature. The competition between these two forces causes the trajectory of the two disclination lines in Fig.~\ref{fig:defectArchitecture}\textit{D} to deviate from regions where $\Delta \theta\left(x,y\right) = \pi/2$. The shape of the disclination lines can then be tuned by changing the magnitude of $\mathbf{f}_B$ and $\mathbf{f}_\gamma$, which vary differently with temperature due to the temperature-sensitive behavior of $\gamma$ and the elastic constants $K, K_2$. As described below, by tuning the disclination shape, we measure $\gamma/\sqrt{KK_2}$ at various temperatures, enabling us to map our experimental observations to simulations.

\begin{figure}[t!]
\includegraphics[width=0.98\columnwidth]{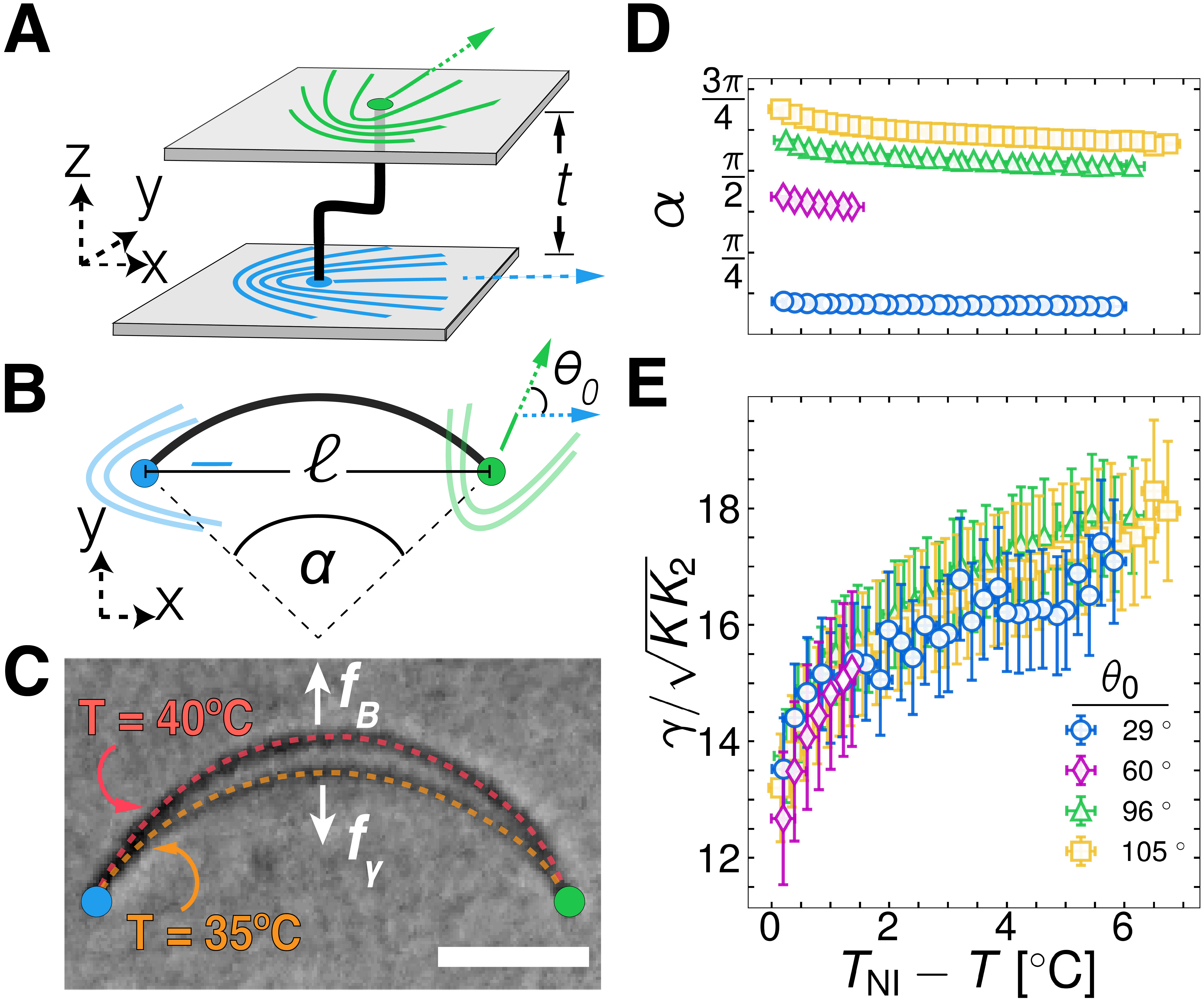}
\caption{ \textbf{Measuring the line tension of a disclination.} ({\textit A}) Diagrams of the geometry used to create arced disclination lines. Initially, $+1/2$ surface defects are imposed as anchoring conditions in a nematic cell. The substrates are then rotated and displaced with respect to one another. ({\textit B}))  When viewed from the top, the defects cores are separated by a distance $\ell$ and their lines of symmetry form an angle $\theta_0$. The two-dimensional projection of the resulting disclination line is a portion of a circular arc with a central angle $\alpha$. ({\textit C})) Bright-field microscopy overlapped images of disclination lines at two different temperatures. The disclination line's curvature depends on the balance of two opposing forces $f_{\gamma}$ and $f_{B}$, whose magnitudes depend on temperature (scale bar: 10$\mu$m, dashed lines are guides for the eye). ({\textit D})) Varying the temperature of the nematic, therefore, results in a change in the defect line curvature and in the angle $\alpha$ as shown in the curves obtained for different values of $\theta_0$. ({\textit E})) When the measured values of $t$, $\ell$, and $\theta_0$ are all accounted for, the implied values of $\gamma/\sqrt{KK_2}$ all collapse onto the same curve as a function of temperature.}
\label{fig:lineTension}
\end{figure}

To illustrate the tuning of disclination shape, we construct LC cells whose confining surfaces are each photo-patterned with a single $+1/2$ defect. The substrates are rotated so that the defects are oriented with respect to one another by an angle $\theta_0$ and are translated so that a horizontal distance $\ell$ separates the defect cores (Fig.~\ref{fig:lineTension}\textit{A, B}). In this design, the 2D projection of the patterns contains a locus of points where $\Delta \theta = \pi/2$ forms a circular arc segment with an opening angle $2\theta_0$ connecting the defects.  Once a cell is filled with 8CB, a disclination line forms to connect the two defect cores (Fig.~\ref{fig:lineTension}\textit{C}).. In general, the line does not follow the arc with opening angle $\theta_0$ due to $\mathbf{f}_\gamma$.  However, along any circular disclination arc that passes between the two surface defect cores, both Eq.~\ref{eq:fexternal} and \ref{eq:ftension} are uniform. Thus, in equilibrium, the disclination still forms an arc, and finding its curvature through force balancing is a simple algebraic problem:
\begin{equation}
    0=\left(\mathbf{f}_B+\mathbf{f}_\gamma\right)\cdot\mathbf{\hat{N}}=
    \frac{\pi K}{\tilde{t}}\left(\frac{\alpha}{2}-\theta_0\right)+\gamma\frac{2}{\ell}\sin\frac{\alpha}{2},
    \label{eq:arcbalance}
\end{equation}
where $\alpha$ is the opening angle of the arc. Rewriting Eq.~\ref{eq:arcbalance} in a dimensionless form, we obtain the following transcendental equation:
    \begin{equation}\label{eq:arcbalancedimless}
    \frac{\alpha}{2}+\tilde{\gamma}\sin\frac{\alpha}{2}=\theta_0,
    \end{equation}
where $\tilde{\gamma}=\frac{2}{\pi}\frac{\gamma}{\sqrt{K K_2}}\frac{t}{\ell}$. As expected, in the limit of vanishing line tension, $\alpha$ tends to $2\theta_0$, where $f_{B}$ vanishes. In the limit of infinite line tension, $\alpha$ tends to zero so that $f_{\gamma}$ vanishes. Line tension's relative importance in determining the defect line's contour is described by the dimensionless parameter $\tilde{\gamma}$.

Equation \ref{eq:arcbalancedimless} captures the effect of line tension in reducing the curvature of an arced disclination line. Rearranging it again, we find that,
\begin{equation}
\frac{\gamma}{\sqrt{KK_2}} = \frac{\pi}{2\kappa t}\left[\theta_{0}-\arcsin\left(\frac{\ell\kappa}{2}\right)\right].
\label{eq:gammaK}
\end{equation}
The equation above links the material parameter $\gamma/\sqrt{KK_2}$ to the deviation of the disclination arc's curvature $\kappa$ from its zero-line tension limit. Thus, the temperature dependence of $\gamma/\sqrt{KK_2}$ can be measured directly in 8CB from the temperature dependence of the line curvature. We track the variation of $\alpha = 2\arcsin\left(\ell\kappa/2\right)$ as a function of temperature across $\theta_0$ ranging from $30^\circ$ to $105^\circ$ (see \textit{Materials and Methods} for details of the image analysis). When a disclination line is formed by an initial $\theta_0 = 105^\circ$, the curvature deep in the nematic phase ($T=35^\circ\rm{C}$) is small (Fig.~\ref{fig:lineTension}\textit{C}). Increasing the temperature towards the nematic-isotropic transition, we observe an increase in $\kappa$ (and hence $\alpha$) since $\mathbf{f}_\gamma$ decreases more rapidly than $\mathbf{f}_{B}$ on heating (Fig.~\ref{fig:lineTension}\textit{D}). Equation \ref{eq:gammaK} is confirmed by the collapse in Fig.~\ref{fig:lineTension}\textit{E} of measurements held at different values of $t$, $\ell$ and $\theta_0$ onto the same curve that only depends on material properties of the LC. This affirms the validity of approximating $\gamma$ with a constant.

Fig.~\ref{fig:lineTension}\textit{E} shows the monotonic temperature dependence of $\gamma/\sqrt{KK_2}$ in a nematic 8CB, ranging approximately between $12$ and $18$. We follow the protocol of Fig.~\ref{fig:lineTension} to also estimate $\gamma/\sqrt{KK_2}$ in our numerical simulations (see \textit{SI Appendix} for details); we analyze arced defect configurations for different values of $t$, $\ell$ and $\theta_0$, and extract $\alpha$ from which we obtain a mean $\gamma/\sqrt{K K_2}=3.3\pm 0.1$. This value is not within the experimental range. However, for every experiment, we can now match a simulation held at the same value of $\tilde{\gamma}=\frac{2}{\pi}\frac{\gamma}{\sqrt{K K_2}}\frac{t}{\ell}$, by compensating for the different values of $\frac{\gamma}{\sqrt{K K_2}}$ with inversely different values of the aspect ratio $\frac{t}{\ell}$. In simulations, we tweak $\tilde{\gamma}$ not with temperature but with aspect ratio.

We now revisit Fig.~\ref{fig:defectArchitecture} and the heart-shaped disclination lines. These are generated using patterns described in detail in the \textit{Materials and Methods}. 
\begin{figure}[t!]
\centering
\includegraphics[width=1\columnwidth]{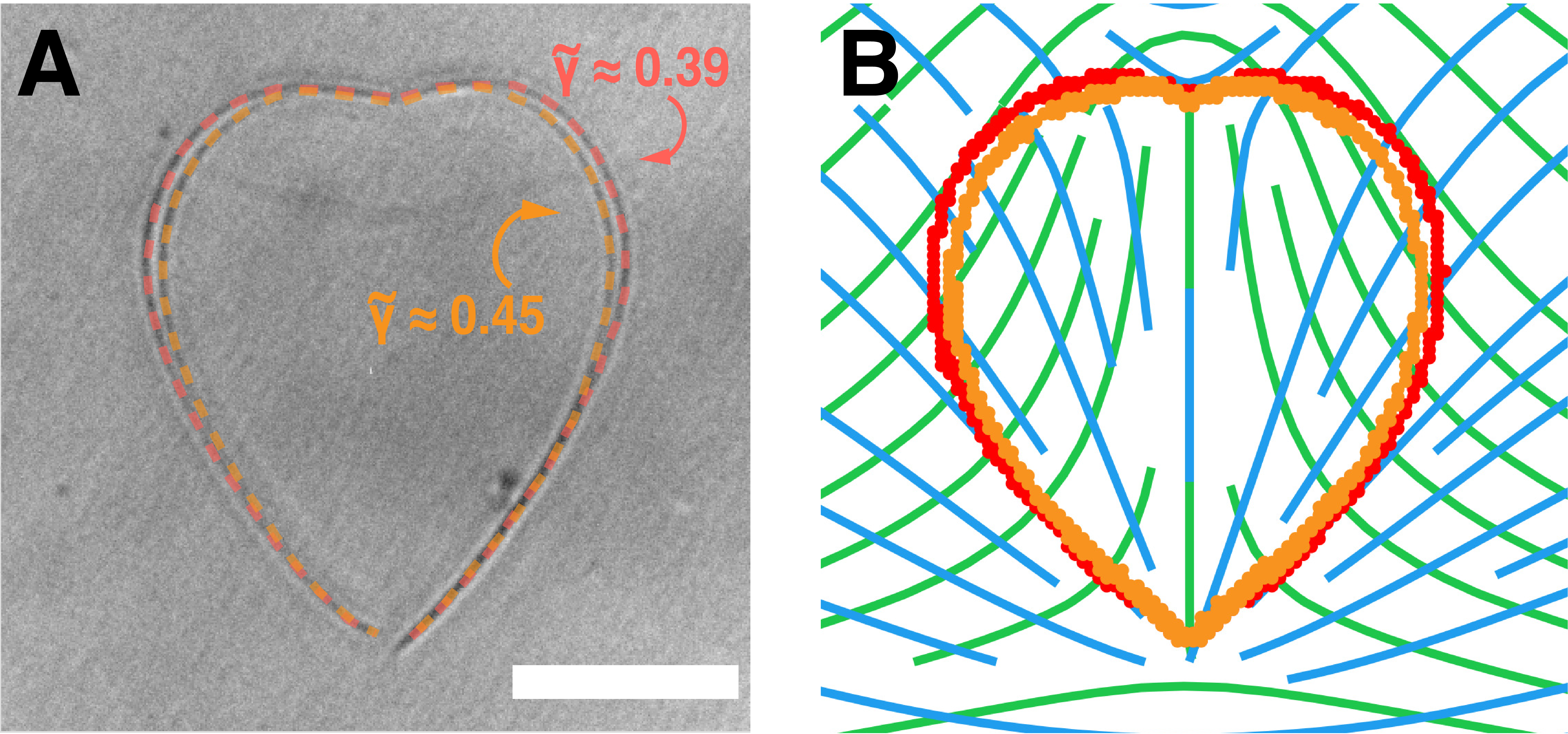}
\caption{ \textbf{Tunable disclination line architecture} 
({\textit A}) Bright-field microscopy image of heart-shaped disclination lines measured for two different temperatures, corresponding to different values of $\tilde{\gamma}$ (scale bar: 25 $\mu$m, dashed lines are guides for the eye). ({\textit B}) Defect configurations obtained in the simulation for two different sets of parameters ${l,t}$, chosen such that the values of $\tilde{\gamma}$ are the same in simulation and experiment. 
The change in the structure of the disclination architecture in both experiment and simulation is captured by $\tilde{\gamma}$.
}
\label{fig:tuning}
\end{figure}
We control the cusps of the heart by the directions of maximum twist around the defects as in Fig.~\ref{fig:defectConnectivity}. When 8CB is cooled by $\approx 6^{\circ}\rm{C}$ from the nematic-isotropic transition, the increase in $\tilde{\gamma}$ constricts the lobes of the heart-shaped disclination lines (Fig.~\ref{fig:tuning}\textit{A}).  We know the value of $\tilde{\gamma}$ at each temperature from the thickness of the cell, the lateral separation between the two surface defects, and Fig.~\ref{fig:lineTension}E. Simulations with the same values of $\tilde{\gamma}$, obtained by changing the values of $t$ and $\ell$, qualitatively capture a similar change in the structure of the disclination architecture Fig.~\ref{fig:tuning}\textit{B}. It is remarkable that despite the experimental uncertainty and the use of different system sizes in the experiment and simulation, the resulting defect configurations for the same values of $\tilde{\gamma}$ are in good agreement. 

\section*{Conclusion}
This work introduces a novel framework for creating and tuning 3D disclination lines in a nematic liquid crystal. When disclination lines are nucleated by surface defects, their connectivity and trajectories are analogous to  current-carrying wires near a current-free surface. Whether or not surface defects may connect to each other can be explained by treating the topologically charged disclination lines as wires that must conserve current. Similarly, substrates imprinted with surface-anchoring conditions exert a Lorentz-like force on the wires, pushing them towards regions where the anchoring conditions on opposing substrates are orthogonal. When the patterns promote wires to curve, they experience an additional force from line tension that decreases the curvature. This force can be tuned in both experiments and simulations by changing a dimensionless parameter, $\tilde{\gamma}$. 

We verified these connectivity principles through a series of experiments. By appropriately designing surface anchoring conditions, we created a three-dimensional structure whose two-dimensional projection resembles a heart. We tuned its shape by varying the temperature and recreated the results using numerical simulations.

Our design principles can be used to interpret similar results observed in recent experiments with disclination lines created by patterned surfaces \cite{nys_nematic_2022,guo_photopatterned_2021,sunami_shape_2018,ouchi_topologically_2019,selingerFrank}. These principles can further be used to construct more complex disclination architecture, advancing the design of tunable 3D liquid crystal-based disclination networks for applications of molecular self-assembly, re-configurable optics, photonic devices, and responsive matter.
Furthermore, we have shown that the equilibrium shape of disclination lines depends on temperature and aspect ratio, opening the door for multi-state systems, switchable by varying the temperature or thickness of the cell.



\section*{Materials and Methods}
\subsection*{Substrate preparation}

Photosensitive material Brilliant Yellow (BY, Sigma-Aldrich) was mixed with n,n-dimethylformamide (DMF) solvent at 1 wt.\% concentration. Glass substrates (Fisher Scientific) were washed in an ultrasonic bath with Hellmanex liquid detergent (Fisherbrand), followed by successive washes in acetone, ethanol, and isopropyl alcohol, and then dried with $\rm{N}_2$ gas. The BY-DMF solution was spin-coated on the substrates at $3000$ RPM for $45$ seconds. After spin-coating, the substrates were baked at $95^\circ\rm{C}$ for $15$ minutes. Spin-coating and baking processes were conducted at a relative humidity of $35\%$ or lower\cite{doi:10.1080/02678292.2016.1247479}. 
\subsection*{Patterned Surface Alignment}
Surface patterns were created using a custom-built photo-patterning setup consisting of a polarized LED source \cite{priimagi} feeding into the side port of a bright-field inverted microscope body (TI Eclipse TE2000). 

Segmented images were generated via a LED-based projector (Sony MPL-C1A) to a peripheral optical path (\textit{SI Appendix} Fig. S1). The projector operates using three time-modulated laser diodes. To match the absorption band of the BY-DMF solution, we use the blue ($\lambda = 445 \rm{ nm}$) diode. Images generated by the projector first pass through two aspheric condenser lenses (with focal lengths $f = 32 \rm{ mm}$, Thorlabs,  ACL50832U) before being expanded with a custom-Keplerian telescope consisting of two convex lenses $f = 100 \rm{ mm}$ (Thorlabs, AC508-100-A-ML) and $f = 200\rm{ mm}$ (Thorlabs, AC508-200-A-ML), respectively. The expanded image passes through a linear polarizer before entering the microscope body. Once inside the body, the image is reflected by a dichroic mirror, picked up by an infinity-corrected tube lens, and collected by a microscope objective (20x, Nikon S Plan Fluor ELWD) that focuses the image onto a BY-coated substrate. Upon irradiation with linearly polarized light, the photosensitive azo-dye molecules orient perpendicularly to the plane of polarization, setting the preferred alignment direction of the nematic director $\mathbf{\hat{n}}$.  

Designed patterns were discretized into pie segments of fixed polarization with opening angle $\pi/16$ and with the cores of defects located at the center.
 
\subsection*{Sample preparation}
 
After photo alignment, patterned regions on substrates were aligned and fixed using epoxy glue (Loctite) to create a liquid crystal cell. After cell assembly, we use spectroscopic reflectometry to measure the cells’ thickness $t$, obtained from the absolute reflectance spectra (Oceanview) fit using custom Matlab code. Cells are subsequently filled with 4’-n-octyl-4-cyano-biphenyl (8CB, Nematel GmbH) liquid crystal, pre-heated into the isotropic phase by capillary flow. After cells are filled, they are  sealed on their ends using UV curable resin (Loon Outdoors UV Clear Fly Finish). 
 
\subsection*{Polarized optical microscopy}
  
We use a Nikon LV 100N Pol upright microscope to image patterned regions with both 20x and 50x objectives.  Samples are placed on a heating stage (Instec HCS302) set to $36^\circ$C to keep 8CB in the nematic phase. Optical microscopy images are captured using a Nikon DS-Ri2 camera.

\subsection*{Analyzing the curvature of disclination line arcs}
Videos of disclination lines are captured using bright-field microscopy and analyzed using ImageJ, TrackPy\cite{allan_daniel_2016_60550}, and custom Python code. The contours of the disclination lines are detected using a Canny edge detection algorithm, binarized, and fit to circles using least squares fitting.
For each frame $i$ of the video, the radius  of curvature $r_i\equiv 1/\kappa$ and center of the best-fit circle $\left(x_i^C,y_i^C\right)$ are determined.  These circles intersect at the defect cores, corresponding to two unique points. To find the positions of these points $(x,y)$, we minimize a cost function $\Xi =$$\sum_i\left(\frac{(x-x_i^c)^2+(y-y_i^c)^2)}{r_i^2}-1\right)^2$, where the sum is over all the frames in the video. The uncertainty of each defect core's position is the cost function's value, and the distance $\ell$ between the defect cores is calculated using the Euclidean distance.
 
\subsection*{Jones matrix calculations}
 For qualitative comparison of numerical and experimental director configurations near nucleated disclination lines, we use Jones calculus to reconstruct the polarized optical microscopy (POM) texture of the director field obtained from minimization of the Landau de-Gennes free-energy. The volume of the numerically-obtained director field is discretized into volume elements (voxels) on a 3D grid, with each point at position $\bm{\rho}$ containing $N$ voxels each of thickness $\Delta$. It is assumed that variation in $\bm{n}$ between successive voxels is small compared to the wavelength of incident light $\lambda$, so that $\lambda \ll 1/ |\nabla\mathbf{\hat{n}}|$.  Each voxel $\nu$ is treated as a uniaxial birefringent optical element, represented by a $2\times2$ Jones matrix $\mathbb{M}_{\nu}$ that depends on both the extraordinary $n^{e}$ and ordinary $n^{o}$ indices of refraction of the LC. Light propagating through a voxel experiences an $n^{e}$ dependant on the polar angle $\theta_{\nu}$ between $\bm{n}$ and the light's  propagation direction $\bm{k}_0$ given by $n^{e}(\theta_{\nu})=n^{o}n^{e}/\sqrt{\left(n^{o}\cos\theta_{\nu}\right)^{2}+\left(n^{e}\sin\theta_{\nu}\right)^{2}}$. We choose $\bm{k}_0=\hat{z}$, so that the plane of polarization is the $x-z$ plane, and write the corresponding Jones matrix as \begin{equation}
 \mathbb{M}_{\nu}\left(\bm{\rho}\right)\equiv\left(\begin{array}{cc}
e^{in_{\nu}^{e}\left(\theta_{\nu}\right)2\pi\frac{\Delta}{\lambda}} & 0\\
0 & e^{in_{\nu}^{o}2\pi\frac{\Delta}{\lambda}}
\end{array}\right).
 \end{equation}
 
Using the 8CB’s $n^{e}$ and $n^{o}$ at the experimental temperature and wavelength $\lambda$, we compute $\mathbb{M}_{\nu}\left(\bm{\rho}\right)$, constructing a single operator $\mathbb{\gamma}\left(\bm{\rho}\right)=\prod_{\nu=1}^{N}\mathbb{R}\left(-\phi_{\nu}\right)\mathbb{M}_{\nu}\left(\bm{\rho}\right)\mathbb{R}\left(\phi_{\nu}\right)$, where $\mathbb{R}\left(\phi_{\nu}\right)\equiv\bigg(\begin{array}{cc}
\cos\phi_\nu & \sin\phi_\nu\\
-\sin\phi_\nu & \cos\phi_\nu
\end{array}\bigg)$ and $\phi_{\nu}$ is the azimuthal component of $\mathbf{\hat{n}}$ in voxel $\nu$. Following \cite{hecht_optics_2017} and \cite{ellis_simulating_2019}, we construct $2\times 2$ Jones matrices for the polarizer $\mathbb{P}$ and analyzer $\mathbb{A}$. Sequential propagation of plane waves $\bm{E_{0}}$ through $\mathbb{P}$,$\mathbb{\gamma}\left(\bm{\rho}\right)$ and $\mathbb{A}$ results in a single Jones vector $\bm{E}_{T}\left(\bm{\rho}\right) = \mathbb{A}\mathbb{\nu}\left(\bm{\rho}\right)\mathbb{P}\bm{E_{0}}$. The calculated POM texture is obtained from  the intensity of light transmitted through each voxel, $I_{T}\left(\bm{\rho}\right) = |\bm{E}_{T}\left(\bm{\rho}\right)|^2$.

\subsection*{Numerical simulations}
The numerical modeling of the nematic liquid crystal is achieved using the lattice-discretized Landau-de Gennes model implemented in open-Qmin\cite{dan_sussman}. The configuration of a nematic liquid crystal is represented by specifying the components of the $Q$-tensor\cite{dan_sussman,Qtensor} which is related to the director $\mathbf{\hat{n}}$ of a uniaxial nematic by $Q_{ij}=\frac{3}{2} S\left( n_i n_j -\frac{1}{3} \delta_{ij}\right)$, where ${i,j} \in \{x,y,z\}$ and $S$ is the degree of uniaxial nematic order. To simulate a thin nematic cell, we consider a three-dimensional box of size $L\times L \times L_z$ with $L \gg L_z$. In the simulation, we use $L=250$ and $L_z$ between $12$ and $21$, expressed in units of the number of lattice sites. Note that the thickness $t=L_z-1$, since anchoring is imposed on top and bottom layers. At every lattice point, we start with a random initial condition for $Q_{ij}$. We impose strong planar anchoring at the top and bottom surfaces by setting the anchoring strength $W=50$ for the two surfaces. We use free boundary conditions on the side surfaces of the simulation box by setting $W=0$. We use the Fast Inertial Relaxation Engine (FIRE) algorithm within open-Qmin\cite{dan_sussman} to minimize the total free energy until the norm of the residual force vector goes below $10^{-8}$ (see \textit{SI Appendix} for details). In the energy-minimized configuration, defects are identified locally as lattice sites where the largest eigenvalue of $Q$ falls below some threshold, typically $0.95S$.\vspace{-5.5mm}
\subsection*{Surface patterns used in experiment and simulation} 
In the experiment and simulation, we impose a planar director field, i.e., the nematic director field takes the form $\mathbf{\hat{n}}=\left(\cos\theta,\sin\theta,0\right)$ at the top and bottom surfaces. In Fig.1 and 4, the surface pattern at the top surface is represented  by
\begin{equation}\label{eq:pattern}
\begin{split}
\theta_{\rm{t}}\left(x, y, z\right) =\frac{1}{2} \left(\tan ^{-1}\frac{y-1}{x}-\tan ^{-1}\frac{y+1}{x} \right. \\
  \left. +\frac{x}{\sqrt{x^2+(y+1)^2}}-\frac{x}{\sqrt{x^2+(y-1)^2}}\right),
\end{split}
\end{equation}
while $\theta_{\rm{b}}\left(x, y, z\right) =-\theta_{\rm{t}}\left(x, y, z\right)$.

\noindent For the $\pm 1/2$ surface defect patterns used in Fig.(2-3) we have $\theta\left(x, y, z\right) =\pm \frac{1}{2} \tan ^{-1}\left(\frac{y}{x}\right)$.

\subsection*{Acknowledgements} 
We greatly acknowledge insights, assistance, and helpful discussions with Charles Rosenblatt, David Dolgitzer and Bastian Pradenas. This research was supported by a grant from the United States-Israel Binational Science Foundation (BSF) no. 2018380. RLL acknowledges support from the NSF (DMR-2104747).

\vspace{4mm}
\bibliographystyle{apsrev4-2}

\bibliography{bibArxiv}

\begin{thebibliography}{37}%
\makeatletter
\providecommand \@ifxundefined [1]{%
 \@ifx{#1\undefined}
}%
\providecommand \@ifnum [1]{%
 \ifnum #1\expandafter \@firstoftwo
 \else \expandafter \@secondoftwo
 \fi
}%
\providecommand \@ifx [1]{%
 \ifx #1\expandafter \@firstoftwo
 \else \expandafter \@secondoftwo
 \fi
}%
\providecommand \natexlab [1]{#1}%
\providecommand \enquote  [1]{``#1''}%
\providecommand \bibnamefont  [1]{#1}%
\providecommand \bibfnamefont [1]{#1}%
\providecommand \citenamefont [1]{#1}%
\providecommand \href@noop [0]{\@secondoftwo}%
\providecommand \href [0]{\begingroup \@sanitize@url \@href}%
\providecommand \@href[1]{\@@startlink{#1}\@@href}%
\providecommand \@@href[1]{\endgroup#1\@@endlink}%
\providecommand \@sanitize@url [0]{\catcode `\\12\catcode `\$12\catcode
  `\&12\catcode `\#12\catcode `\^12\catcode `\_12\catcode `\%12\relax}%
\providecommand \@@startlink[1]{}%
\providecommand \@@endlink[0]{}%
\providecommand \url  [0]{\begingroup\@sanitize@url \@url }%
\providecommand \@url [1]{\endgroup\@href {#1}{\urlprefix }}%
\providecommand \urlprefix  [0]{URL }%
\providecommand \Eprint [0]{\href }%
\providecommand \doibase [0]{https://doi.org/}%
\providecommand \selectlanguage [0]{\@gobble}%
\providecommand \bibinfo  [0]{\@secondoftwo}%
\providecommand \bibfield  [0]{\@secondoftwo}%
\providecommand \translation [1]{[#1]}%
\providecommand \BibitemOpen [0]{}%
\providecommand \bibitemStop [0]{}%
\providecommand \bibitemNoStop [0]{.\EOS\space}%
\providecommand \EOS [0]{\spacefactor3000\relax}%
\providecommand \BibitemShut  [1]{\csname bibitem#1\endcsname}%
\let\auto@bib@innerbib\@empty
\bibitem [{\citenamefont {Kosterlitz}\ and\ \citenamefont
  {Thouless}(1973)}]{kosterlitz_ordering_1973}%
  \BibitemOpen
  \bibfield  {author} {\bibinfo {author} {\bibfnamefont {J.~M.}\ \bibnamefont
  {Kosterlitz}}\ and\ \bibinfo {author} {\bibfnamefont {D.~J.}\ \bibnamefont
  {Thouless}},\ }\href {https://doi.org/10.1088/0022-3719/6/7/010} {\bibfield
  {journal} {\bibinfo  {journal} {Journal of Physics C: Solid State Physics}\
  }\textbf {\bibinfo {volume} {6}},\ \bibinfo {pages} {1181} (\bibinfo {year}
  {1973})}\BibitemShut {NoStop}%
\bibitem [{\citenamefont {Maroudas-Sacks}\ \emph {et~al.}(2021)\citenamefont
  {Maroudas-Sacks}, \citenamefont {Garion}, \citenamefont {Shani-Zerbib},
  \citenamefont {Livshits}, \citenamefont {Braun},\ and\ \citenamefont
  {Keren}}]{maroudas-sacks_topological_2021}%
  \BibitemOpen
  \bibfield  {author} {\bibinfo {author} {\bibfnamefont {Y.}~\bibnamefont
  {Maroudas-Sacks}}, \bibinfo {author} {\bibfnamefont {L.}~\bibnamefont
  {Garion}}, \bibinfo {author} {\bibfnamefont {L.}~\bibnamefont
  {Shani-Zerbib}}, \bibinfo {author} {\bibfnamefont {A.}~\bibnamefont
  {Livshits}}, \bibinfo {author} {\bibfnamefont {E.}~\bibnamefont {Braun}},\
  and\ \bibinfo {author} {\bibfnamefont {K.}~\bibnamefont {Keren}},\ }\href
  {https://doi.org/10.1038/s41567-020-01083-1} {\bibfield  {journal} {\bibinfo
  {journal} {Nature Physics}\ }\textbf {\bibinfo {volume} {17}},\ \bibinfo
  {pages} {251} (\bibinfo {year} {2021})}\BibitemShut {NoStop}%
\bibitem [{\citenamefont {Brasselet}(2012)}]{brasselet_tunable_2012}%
  \BibitemOpen
  \bibfield  {author} {\bibinfo {author} {\bibfnamefont {E.}~\bibnamefont
  {Brasselet}},\ }\href {https://doi.org/10.1103/PhysRevLett.108.087801}
  {\bibfield  {journal} {\bibinfo  {journal} {Physical Review Letters}\
  }\textbf {\bibinfo {volume} {108}},\ \bibinfo {pages} {087801} (\bibinfo
  {year} {2012})}\BibitemShut {NoStop}%
\bibitem [{\citenamefont {Meng}\ \emph {et~al.}(2023)\citenamefont {Meng},
  \citenamefont {Wu},\ and\ \citenamefont {Smalyukh}}]{meng_topological_2023}%
  \BibitemOpen
  \bibfield  {author} {\bibinfo {author} {\bibfnamefont {C.}~\bibnamefont
  {Meng}}, \bibinfo {author} {\bibfnamefont {J.-S.}\ \bibnamefont {Wu}},\ and\
  \bibinfo {author} {\bibfnamefont {I.~I.}\ \bibnamefont {Smalyukh}},\ }\href
  {https://doi.org/10.1038/s41563-022-01414-y} {\bibfield  {journal} {\bibinfo
  {journal} {Nature Materials}\ }\textbf {\bibinfo {volume} {22}},\ \bibinfo
  {pages} {64} (\bibinfo {year} {2023})}\BibitemShut {NoStop}%
\bibitem [{\citenamefont {Blanc}\ \emph {et~al.}(2013)\citenamefont {Blanc},
  \citenamefont {Coursault},\ and\ \citenamefont
  {Lacaze}}]{blanc_ordering_2013}%
  \BibitemOpen
  \bibfield  {author} {\bibinfo {author} {\bibfnamefont {C.}~\bibnamefont
  {Blanc}}, \bibinfo {author} {\bibfnamefont {D.}~\bibnamefont {Coursault}},\
  and\ \bibinfo {author} {\bibfnamefont {E.}~\bibnamefont {Lacaze}},\ }\href
  {https://doi.org/10.1080/21680396.2013.818515} {\bibfield  {journal}
  {\bibinfo  {journal} {Liquid Crystals Reviews}\ }\textbf {\bibinfo {volume}
  {1}},\ \bibinfo {pages} {83} (\bibinfo {year} {2013})}\BibitemShut {NoStop}%
\bibitem [{\citenamefont {Luo}\ \emph {et~al.}(2018)\citenamefont {Luo},
  \citenamefont {Beller}, \citenamefont {Boniello}, \citenamefont {Serra},\
  and\ \citenamefont {Stebe}}]{luo_tunable_2018}%
  \BibitemOpen
  \bibfield  {author} {\bibinfo {author} {\bibfnamefont {Y.}~\bibnamefont
  {Luo}}, \bibinfo {author} {\bibfnamefont {D.~A.}\ \bibnamefont {Beller}},
  \bibinfo {author} {\bibfnamefont {G.}~\bibnamefont {Boniello}}, \bibinfo
  {author} {\bibfnamefont {F.}~\bibnamefont {Serra}},\ and\ \bibinfo {author}
  {\bibfnamefont {K.~J.}\ \bibnamefont {Stebe}},\ }\href
  {https://doi.org/10.1038/s41467-018-06054-y} {\bibfield  {journal} {\bibinfo
  {journal} {Nature Communications}\ }\textbf {\bibinfo {volume} {9}},\
  \bibinfo {pages} {3841} (\bibinfo {year} {2018})}\BibitemShut {NoStop}%
\bibitem [{\citenamefont {Coles}\ and\ \citenamefont
  {Pivnenko}(2005)}]{coles_liquid_2005}%
  \BibitemOpen
  \bibfield  {author} {\bibinfo {author} {\bibfnamefont {H.~J.}\ \bibnamefont
  {Coles}}\ and\ \bibinfo {author} {\bibfnamefont {M.~N.}\ \bibnamefont
  {Pivnenko}},\ }\href {https://doi.org/10.1038/nature03932} {\bibfield
  {journal} {\bibinfo  {journal} {Nature}\ }\textbf {\bibinfo {volume} {436}},\
  \bibinfo {pages} {997} (\bibinfo {year} {2005})}\BibitemShut {NoStop}%
\bibitem [{\citenamefont {Sengupta}\ \emph {et~al.}(2013)\citenamefont
  {Sengupta}, \citenamefont {Bahr},\ and\ \citenamefont
  {Herminghaus}}]{sengupta_topological_2013}%
  \BibitemOpen
  \bibfield  {author} {\bibinfo {author} {\bibfnamefont {A.}~\bibnamefont
  {Sengupta}}, \bibinfo {author} {\bibfnamefont {C.}~\bibnamefont {Bahr}},\
  and\ \bibinfo {author} {\bibfnamefont {S.}~\bibnamefont {Herminghaus}},\
  }\href {https://doi.org/10.1039/c3sm50677k} {\bibfield  {journal} {\bibinfo
  {journal} {Soft Matter}\ }\textbf {\bibinfo {volume} {9}},\ \bibinfo {pages}
  {7251} (\bibinfo {year} {2013})}\BibitemShut {NoStop}%
\bibitem [{\citenamefont {Sandford~O’Neill}\ \emph
  {et~al.}(2020)\citenamefont {Sandford~O’Neill}, \citenamefont {Salter},
  \citenamefont {Booth}, \citenamefont {Elston},\ and\ \citenamefont
  {Morris}}]{sandford_oneill_electrically-tunable_2020}%
  \BibitemOpen
  \bibfield  {author} {\bibinfo {author} {\bibfnamefont {J.~J.}\ \bibnamefont
  {Sandford~O’Neill}}, \bibinfo {author} {\bibfnamefont {P.~S.}\ \bibnamefont
  {Salter}}, \bibinfo {author} {\bibfnamefont {M.~J.}\ \bibnamefont {Booth}},
  \bibinfo {author} {\bibfnamefont {S.~J.}\ \bibnamefont {Elston}},\ and\
  \bibinfo {author} {\bibfnamefont {S.~M.}\ \bibnamefont {Morris}},\ }\href
  {https://doi.org/10.1038/s41467-020-16059-1} {\bibfield  {journal} {\bibinfo
  {journal} {Nature Communications}\ }\textbf {\bibinfo {volume} {11}},\
  \bibinfo {pages} {2203} (\bibinfo {year} {2020})}\BibitemShut {NoStop}%
\bibitem [{\citenamefont {Murray}\ \emph {et~al.}(2014)\citenamefont {Murray},
  \citenamefont {Pelcovits},\ and\ \citenamefont {Rosenblatt}}]{chuck}%
  \BibitemOpen
  \bibfield  {author} {\bibinfo {author} {\bibfnamefont {B.~S.}\ \bibnamefont
  {Murray}}, \bibinfo {author} {\bibfnamefont {R.~A.}\ \bibnamefont
  {Pelcovits}},\ and\ \bibinfo {author} {\bibfnamefont {C.}~\bibnamefont
  {Rosenblatt}},\ }\href {https://doi.org/10.1103/PhysRevE.90.052501}
  {\bibfield  {journal} {\bibinfo  {journal} {Physical Review E}\ }\textbf
  {\bibinfo {volume} {90}},\ \bibinfo {pages} {052501} (\bibinfo {year}
  {2014})}\BibitemShut {NoStop}%
\bibitem [{\citenamefont {Harkai}\ \emph {et~al.}(2020)\citenamefont {Harkai},
  \citenamefont {Murray}, \citenamefont {Rosenblatt},\ and\ \citenamefont
  {Kralj}}]{harkai_electric_2020}%
  \BibitemOpen
  \bibfield  {author} {\bibinfo {author} {\bibfnamefont {S.}~\bibnamefont
  {Harkai}}, \bibinfo {author} {\bibfnamefont {B.~S.}\ \bibnamefont {Murray}},
  \bibinfo {author} {\bibfnamefont {C.}~\bibnamefont {Rosenblatt}},\ and\
  \bibinfo {author} {\bibfnamefont {S.}~\bibnamefont {Kralj}},\ }\href
  {https://doi.org/10.1103/PhysRevResearch.2.013176} {\bibfield  {journal}
  {\bibinfo  {journal} {Physical Review Research}\ }\textbf {\bibinfo {volume}
  {2}},\ \bibinfo {pages} {013176} (\bibinfo {year} {2020})}\BibitemShut
  {NoStop}%
\bibitem [{\citenamefont {Wang}\ \emph
  {et~al.}(2017{\natexlab{a}})\citenamefont {Wang}, \citenamefont {Li},\ and\
  \citenamefont {Yokoyama}}]{wang_artificial_2017}%
  \BibitemOpen
  \bibfield  {author} {\bibinfo {author} {\bibfnamefont {M.}~\bibnamefont
  {Wang}}, \bibinfo {author} {\bibfnamefont {Y.}~\bibnamefont {Li}},\ and\
  \bibinfo {author} {\bibfnamefont {H.}~\bibnamefont {Yokoyama}},\ }\href
  {https://doi.org/10.1038/s41467-017-00548-x} {\bibfield  {journal} {\bibinfo
  {journal} {Nature Communications}\ }\textbf {\bibinfo {volume} {8}},\
  \bibinfo {pages} {388} (\bibinfo {year} {2017}{\natexlab{a}})}\BibitemShut
  {NoStop}%
\bibitem [{\citenamefont {Ouchi}\ \emph {et~al.}(2019)\citenamefont {Ouchi},
  \citenamefont {Imamura}, \citenamefont {Sunami}, \citenamefont {Yoshida},\
  and\ \citenamefont {Ozaki}}]{ouchi_topologically_2019}%
  \BibitemOpen
  \bibfield  {author} {\bibinfo {author} {\bibfnamefont {T.}~\bibnamefont
  {Ouchi}}, \bibinfo {author} {\bibfnamefont {K.}~\bibnamefont {Imamura}},
  \bibinfo {author} {\bibfnamefont {K.}~\bibnamefont {Sunami}}, \bibinfo
  {author} {\bibfnamefont {H.}~\bibnamefont {Yoshida}},\ and\ \bibinfo {author}
  {\bibfnamefont {M.}~\bibnamefont {Ozaki}},\ }\href
  {https://doi.org/10.1103/PhysRevLett.123.097801} {\bibfield  {journal}
  {\bibinfo  {journal} {Physical Review Letters}\ }\textbf {\bibinfo {volume}
  {123}},\ \bibinfo {pages} {097801} (\bibinfo {year} {2019})}\BibitemShut
  {NoStop}%
\bibitem [{\citenamefont {Sunami}\ \emph {et~al.}(2018)\citenamefont {Sunami},
  \citenamefont {Imamura}, \citenamefont {Ouchi}, \citenamefont {Yoshida},\
  and\ \citenamefont {Ozaki}}]{sunami_shape_2018}%
  \BibitemOpen
  \bibfield  {author} {\bibinfo {author} {\bibfnamefont {K.}~\bibnamefont
  {Sunami}}, \bibinfo {author} {\bibfnamefont {K.}~\bibnamefont {Imamura}},
  \bibinfo {author} {\bibfnamefont {T.}~\bibnamefont {Ouchi}}, \bibinfo
  {author} {\bibfnamefont {H.}~\bibnamefont {Yoshida}},\ and\ \bibinfo {author}
  {\bibfnamefont {M.}~\bibnamefont {Ozaki}},\ }\href
  {https://doi.org/10.1103/PhysRevE.97.020701} {\bibfield  {journal} {\bibinfo
  {journal} {Physical Review E}\ }\textbf {\bibinfo {volume} {97}},\ \bibinfo
  {pages} {020701} (\bibinfo {year} {2018})}\BibitemShut {NoStop}%
\bibitem [{\citenamefont {Guo}\ \emph {et~al.}(2021)\citenamefont {Guo},
  \citenamefont {Jiang}, \citenamefont {Afghah}, \citenamefont {Peng},
  \citenamefont {Selinger}, \citenamefont {Lavrentovich},\ and\ \citenamefont
  {Wei}}]{guo_photopatterned_2021}%
  \BibitemOpen
  \bibfield  {author} {\bibinfo {author} {\bibfnamefont {Y.}~\bibnamefont
  {Guo}}, \bibinfo {author} {\bibfnamefont {M.}~\bibnamefont {Jiang}}, \bibinfo
  {author} {\bibfnamefont {S.}~\bibnamefont {Afghah}}, \bibinfo {author}
  {\bibfnamefont {C.}~\bibnamefont {Peng}}, \bibinfo {author} {\bibfnamefont
  {R.~L.~B.}\ \bibnamefont {Selinger}}, \bibinfo {author} {\bibfnamefont
  {O.~D.}\ \bibnamefont {Lavrentovich}},\ and\ \bibinfo {author} {\bibfnamefont
  {Q.}~\bibnamefont {Wei}},\ }\href {https://doi.org/10.1002/adom.202170063}
  {\bibfield  {journal} {\bibinfo  {journal} {Advanced Optical Materials}\
  }\textbf {\bibinfo {volume} {9}},\ \bibinfo {pages} {2170063} (\bibinfo
  {year} {2021})}\BibitemShut {NoStop}%
\bibitem [{\citenamefont {Yoshida}\ \emph {et~al.}(2015)\citenamefont
  {Yoshida}, \citenamefont {Asakura}, \citenamefont {Fukuda},\ and\
  \citenamefont {Ozaki}}]{yoshida_three-dimensional_2015}%
  \BibitemOpen
  \bibfield  {author} {\bibinfo {author} {\bibfnamefont {H.}~\bibnamefont
  {Yoshida}}, \bibinfo {author} {\bibfnamefont {K.}~\bibnamefont {Asakura}},
  \bibinfo {author} {\bibfnamefont {J.}~\bibnamefont {Fukuda}},\ and\ \bibinfo
  {author} {\bibfnamefont {M.}~\bibnamefont {Ozaki}},\ }\href
  {https://doi.org/10.1038/ncomms8180} {\bibfield  {journal} {\bibinfo
  {journal} {Nature Communications}\ }\textbf {\bibinfo {volume} {6}},\
  \bibinfo {pages} {7180} (\bibinfo {year} {2015})}\BibitemShut {NoStop}%
\bibitem [{\citenamefont {Nys}\ \emph {et~al.}(2022)\citenamefont {Nys},
  \citenamefont {Berteloot}, \citenamefont {Beeckman},\ and\ \citenamefont
  {Neyts}}]{nys_nematic_2022}%
  \BibitemOpen
  \bibfield  {author} {\bibinfo {author} {\bibfnamefont {I.}~\bibnamefont
  {Nys}}, \bibinfo {author} {\bibfnamefont {B.}~\bibnamefont {Berteloot}},
  \bibinfo {author} {\bibfnamefont {J.}~\bibnamefont {Beeckman}},\ and\
  \bibinfo {author} {\bibfnamefont {K.}~\bibnamefont {Neyts}},\ }\href
  {https://doi.org/10.1002/adom.202101626} {\bibfield  {journal} {\bibinfo
  {journal} {Advanced Optical Materials}\ }\textbf {\bibinfo {volume} {10}},\
  \bibinfo {pages} {2101626} (\bibinfo {year} {2022})}\BibitemShut {NoStop}%
\bibitem [{\citenamefont {Nys}\ \emph {et~al.}(2018)\citenamefont {Nys},
  \citenamefont {Chen}, \citenamefont {Beeckman},\ and\ \citenamefont
  {Neyts}}]{nys_periodic_2018}%
  \BibitemOpen
  \bibfield  {author} {\bibinfo {author} {\bibfnamefont {I.}~\bibnamefont
  {Nys}}, \bibinfo {author} {\bibfnamefont {K.}~\bibnamefont {Chen}}, \bibinfo
  {author} {\bibfnamefont {J.}~\bibnamefont {Beeckman}},\ and\ \bibinfo
  {author} {\bibfnamefont {K.}~\bibnamefont {Neyts}},\ }\href
  {https://doi.org/10.1002/adom.201701163} {\bibfield  {journal} {\bibinfo
  {journal} {Advanced Optical Materials}\ }\textbf {\bibinfo {volume} {6}},\
  \bibinfo {pages} {1701163} (\bibinfo {year} {2018})}\BibitemShut {NoStop}%
\bibitem [{\citenamefont {Berteloot}\ \emph {et~al.}(2020)\citenamefont
  {Berteloot}, \citenamefont {Nys}, \citenamefont {Poy}, \citenamefont
  {Beeckman},\ and\ \citenamefont {Neyts}}]{berteloot_ring-shaped_2020}%
  \BibitemOpen
  \bibfield  {author} {\bibinfo {author} {\bibfnamefont {B.}~\bibnamefont
  {Berteloot}}, \bibinfo {author} {\bibfnamefont {I.}~\bibnamefont {Nys}},
  \bibinfo {author} {\bibfnamefont {G.}~\bibnamefont {Poy}}, \bibinfo {author}
  {\bibfnamefont {J.}~\bibnamefont {Beeckman}},\ and\ \bibinfo {author}
  {\bibfnamefont {K.}~\bibnamefont {Neyts}},\ }\href
  {https://doi.org/10.1039/D0SM00308E} {\bibfield  {journal} {\bibinfo
  {journal} {Soft Matter}\ }\textbf {\bibinfo {volume} {16}},\ \bibinfo {pages}
  {4999} (\bibinfo {year} {2020})}\BibitemShut {NoStop}%
\bibitem [{\citenamefont {Jiang}\ \emph {et~al.}(2022)\citenamefont {Jiang},
  \citenamefont {Ranabhat}, \citenamefont {Wang}, \citenamefont {Rich},
  \citenamefont {Zhang},\ and\ \citenamefont {Peng}}]{jiang_active_2022}%
  \BibitemOpen
  \bibfield  {author} {\bibinfo {author} {\bibfnamefont {J.}~\bibnamefont
  {Jiang}}, \bibinfo {author} {\bibfnamefont {K.}~\bibnamefont {Ranabhat}},
  \bibinfo {author} {\bibfnamefont {X.}~\bibnamefont {Wang}}, \bibinfo {author}
  {\bibfnamefont {H.}~\bibnamefont {Rich}}, \bibinfo {author} {\bibfnamefont
  {R.}~\bibnamefont {Zhang}},\ and\ \bibinfo {author} {\bibfnamefont
  {C.}~\bibnamefont {Peng}},\ }\href {https://doi.org/10.1073/pnas.2122226119}
  {\bibfield  {journal} {\bibinfo  {journal} {Proceedings of the National
  Academy of Sciences}\ }\textbf {\bibinfo {volume} {119}},\ \bibinfo {pages}
  {e2122226119} (\bibinfo {year} {2022})}\BibitemShut {NoStop}%
\bibitem [{\citenamefont {Gennes}\ and\ \citenamefont
  {Prost}(2013)}]{deGennesBook}%
  \BibitemOpen
  \bibfield  {author} {\bibinfo {author} {\bibfnamefont {P.-G.~d.}\
  \bibnamefont {Gennes}}\ and\ \bibinfo {author} {\bibfnamefont
  {J.}~\bibnamefont {Prost}},\ }\href@noop {} {\emph {\bibinfo {title} {The
  Physics of Liquid Crystals}}},\ \bibinfo {edition} {2nd}\ ed.,\ \bibinfo
  {series} {International series of monographs on physics}\ No.~\bibinfo
  {number} {83}\ (\bibinfo  {publisher} {Clarendon Press},\ \bibinfo {address}
  {Oxford},\ \bibinfo {year} {2013})\BibitemShut {NoStop}%
\bibitem [{\citenamefont {Love}(1888)}]{Love1888}%
  \BibitemOpen
  \bibfield  {author} {\bibinfo {author} {\bibfnamefont {E.}~\bibnamefont
  {Love}},\ }\href {https://doi.org/10.1098/rsta.1888.0016} {\bibfield
  {journal} {\bibinfo  {journal} {Philosophical Transactions of the Royal
  Society of London. (A.)}\ }\textbf {\bibinfo {volume} {179}},\ \bibinfo
  {pages} {491} (\bibinfo {year} {1888})}\BibitemShut {NoStop}%
\bibitem [{\citenamefont {Oswald}\ \emph {et~al.}(2004)\citenamefont {Oswald},
  \citenamefont {Baudry},\ and\ \citenamefont {Rondepierre}}]{elasticConsts}%
  \BibitemOpen
  \bibfield  {author} {\bibinfo {author} {\bibfnamefont {P.}~\bibnamefont
  {Oswald}}, \bibinfo {author} {\bibfnamefont {J.}~\bibnamefont {Baudry}},\
  and\ \bibinfo {author} {\bibfnamefont {T.}~\bibnamefont {Rondepierre}},\
  }\href {https://doi.org/10.1103/PhysRevE.70.041702} {\bibfield  {journal}
  {\bibinfo  {journal} {Phys. Rev. E}\ }\textbf {\bibinfo {volume} {70}},\
  \bibinfo {pages} {041702} (\bibinfo {year} {2004})}\BibitemShut {NoStop}%
\bibitem [{\citenamefont {Schopohl}\ and\ \citenamefont
  {Sluckin}(1987)}]{Schopohl1987}%
  \BibitemOpen
  \bibfield  {author} {\bibinfo {author} {\bibfnamefont {N.}~\bibnamefont
  {Schopohl}}\ and\ \bibinfo {author} {\bibfnamefont {T.~J.}\ \bibnamefont
  {Sluckin}},\ }\href {https://doi.org/10.1103/PhysRevLett.59.2582} {\bibfield
  {journal} {\bibinfo  {journal} {Phys. Rev. Lett.}\ }\textbf {\bibinfo
  {volume} {59}},\ \bibinfo {pages} {2582} (\bibinfo {year}
  {1987})}\BibitemShut {NoStop}%
\bibitem [{\citenamefont {Mermin}(1979)}]{mermin_topological_1979}%
  \BibitemOpen
  \bibfield  {author} {\bibinfo {author} {\bibfnamefont {N.~D.}\ \bibnamefont
  {Mermin}},\ }\href {https://doi.org/10.1103/RevModPhys.51.591} {\bibfield
  {journal} {\bibinfo  {journal} {Reviews of Modern Physics}\ }\textbf
  {\bibinfo {volume} {51}},\ \bibinfo {pages} {591} (\bibinfo {year}
  {1979})}\BibitemShut {NoStop}%
\bibitem [{\citenamefont {Chigrinov}\ \emph {et~al.}(2008)\citenamefont
  {Chigrinov}, \citenamefont {Kozenkov},\ and\ \citenamefont
  {Kwok}}]{chigrinov_photoalignment_2008}%
  \BibitemOpen
  \bibfield  {author} {\bibinfo {author} {\bibfnamefont {V.~G.}\ \bibnamefont
  {Chigrinov}}, \bibinfo {author} {\bibfnamefont {V.~M.}\ \bibnamefont
  {Kozenkov}},\ and\ \bibinfo {author} {\bibfnamefont {H.-S.}\ \bibnamefont
  {Kwok}},\ }\href@noop {} {\emph {\bibinfo {title} {Photoalignment of liquid
  crystalline materials: physics and applications}}},\ Wiley {SID} series in
  display technology\ (\bibinfo  {publisher} {Wiley},\ \bibinfo {address}
  {Chichester, England ; Hoboken, NJ},\ \bibinfo {year} {2008})\ \bibinfo
  {note} {oCLC: ocn225429570}\BibitemShut {NoStop}%
\bibitem [{\citenamefont {Folwill}\ \emph {et~al.}(2021)\citenamefont
  {Folwill}, \citenamefont {Zeitouny}, \citenamefont {Lall},\ and\
  \citenamefont {Zappe}}]{folwill_practical_2021}%
  \BibitemOpen
  \bibfield  {author} {\bibinfo {author} {\bibfnamefont {Y.}~\bibnamefont
  {Folwill}}, \bibinfo {author} {\bibfnamefont {Z.}~\bibnamefont {Zeitouny}},
  \bibinfo {author} {\bibfnamefont {J.}~\bibnamefont {Lall}},\ and\ \bibinfo
  {author} {\bibfnamefont {H.}~\bibnamefont {Zappe}},\ }\href
  {https://doi.org/10.1080/02678292.2020.1825842} {\bibfield  {journal}
  {\bibinfo  {journal} {Liquid Crystals}\ }\textbf {\bibinfo {volume} {48}},\
  \bibinfo {pages} {862} (\bibinfo {year} {2021})}\BibitemShut {NoStop}%
\bibitem [{\citenamefont {Wani}\ \emph {et~al.}(2017)\citenamefont {Wani},
  \citenamefont {Wasylczyk},\ and\ \citenamefont {Priimagi}}]{priimagi}%
  \BibitemOpen
  \bibfield  {author} {\bibinfo {author} {\bibfnamefont {O.~M.}\ \bibnamefont
  {Wani}}, \bibinfo {author} {\bibfnamefont {P.}~\bibnamefont {Wasylczyk}},\
  and\ \bibinfo {author} {\bibfnamefont {A.}~\bibnamefont {Priimagi}},\
  }\href@noop {} {\bibfield  {journal} {\bibinfo  {journal} {Adv. Opt. Mater.}\
  }\textbf {\bibinfo {volume} {6}},\ \bibinfo {pages} {1700949} (\bibinfo
  {year} {2017})}\BibitemShut {NoStop}%
\bibitem [{\citenamefont {Long}\ \emph {et~al.}(2022)\citenamefont {Long},
  \citenamefont {Deutsch}, \citenamefont {Angelo}, \citenamefont {Culbreath},
  \citenamefont {Yokoyama}, \citenamefont {Selinger},\ and\ \citenamefont
  {Selinger}}]{selingerFrank}%
  \BibitemOpen
  \bibfield  {author} {\bibinfo {author} {\bibfnamefont {C.}~\bibnamefont
  {Long}}, \bibinfo {author} {\bibfnamefont {M.~J.}\ \bibnamefont {Deutsch}},
  \bibinfo {author} {\bibfnamefont {J.}~\bibnamefont {Angelo}}, \bibinfo
  {author} {\bibfnamefont {C.}~\bibnamefont {Culbreath}}, \bibinfo {author}
  {\bibfnamefont {H.}~\bibnamefont {Yokoyama}}, \bibinfo {author}
  {\bibfnamefont {J.~V.}\ \bibnamefont {Selinger}},\ and\ \bibinfo {author}
  {\bibfnamefont {R.~L.~B.}\ \bibnamefont {Selinger}},\ }\href
  {https://doi.org/10.48550/ARXIV.2212.01316} {\bibinfo {title} {Frank-read
  mechanism in nematic liquid crystals}} (\bibinfo {year} {2022})\BibitemShut
  {NoStop}%
\bibitem [{\citenamefont {Wang}\ \emph
  {et~al.}(2017{\natexlab{b}})\citenamefont {Wang}, \citenamefont {McGinty},
  \citenamefont {West}, \citenamefont {Bryant}, \citenamefont {Finnemeyer},
  \citenamefont {Reich}, \citenamefont {Berry}, \citenamefont {Clark},
  \citenamefont {Yaroshchuk},\ and\ \citenamefont
  {Bos}}]{doi:10.1080/02678292.2016.1247479}%
  \BibitemOpen
  \bibfield  {author} {\bibinfo {author} {\bibfnamefont {J.}~\bibnamefont
  {Wang}}, \bibinfo {author} {\bibfnamefont {C.}~\bibnamefont {McGinty}},
  \bibinfo {author} {\bibfnamefont {J.}~\bibnamefont {West}}, \bibinfo {author}
  {\bibfnamefont {D.}~\bibnamefont {Bryant}}, \bibinfo {author} {\bibfnamefont
  {V.}~\bibnamefont {Finnemeyer}}, \bibinfo {author} {\bibfnamefont
  {R.}~\bibnamefont {Reich}}, \bibinfo {author} {\bibfnamefont
  {S.}~\bibnamefont {Berry}}, \bibinfo {author} {\bibfnamefont
  {H.}~\bibnamefont {Clark}}, \bibinfo {author} {\bibfnamefont
  {O.}~\bibnamefont {Yaroshchuk}},\ and\ \bibinfo {author} {\bibfnamefont
  {P.}~\bibnamefont {Bos}},\ }\href
  {https://doi.org/10.1080/02678292.2016.1247479} {\bibfield  {journal}
  {\bibinfo  {journal} {Liquid Crystals}\ }\textbf {\bibinfo {volume} {44}},\
  \bibinfo {pages} {863} (\bibinfo {year} {2017}{\natexlab{b}})},\ \Eprint
  {https://arxiv.org/abs/https://doi.org/10.1080/02678292.2016.1247479}
  {https://doi.org/10.1080/02678292.2016.1247479} \BibitemShut {NoStop}%
\bibitem [{\citenamefont {Allan}\ \emph {et~al.}(2016)\citenamefont {Allan},
  \citenamefont {Caswell}, \citenamefont {Keim},\ and\ \citenamefont {van~der
  Wel}}]{allan_daniel_2016_60550}%
  \BibitemOpen
  \bibfield  {author} {\bibinfo {author} {\bibfnamefont {D.}~\bibnamefont
  {Allan}}, \bibinfo {author} {\bibfnamefont {T.}~\bibnamefont {Caswell}},
  \bibinfo {author} {\bibfnamefont {N.}~\bibnamefont {Keim}},\ and\ \bibinfo
  {author} {\bibfnamefont {C.}~\bibnamefont {van~der Wel}},\ }\href
  {https://doi.org/10.5281/zenodo.60550} {\bibinfo {title} {trackpy: Trackpy
  v0.3.2}} (\bibinfo {year} {2016})\BibitemShut {NoStop}%
\bibitem [{\citenamefont {Hecht}(2017)}]{hecht_optics_2017}%
  \BibitemOpen
  \bibfield  {author} {\bibinfo {author} {\bibfnamefont {E.}~\bibnamefont
  {Hecht}},\ }\href@noop {} {\emph {\bibinfo {title} {Optics}}},\ \bibinfo
  {edition} {5th}\ ed.\ (\bibinfo  {publisher} {Pearson Education, Inc},\
  \bibinfo {address} {Boston},\ \bibinfo {year} {2017})\BibitemShut {NoStop}%
\bibitem [{\citenamefont {Ellis}\ \emph {et~al.}(2019)\citenamefont {Ellis},
  \citenamefont {Pairam},\ and\ \citenamefont
  {Fernández-Nieves}}]{ellis_simulating_2019}%
  \BibitemOpen
  \bibfield  {author} {\bibinfo {author} {\bibfnamefont {P.~W.}\ \bibnamefont
  {Ellis}}, \bibinfo {author} {\bibfnamefont {E.}~\bibnamefont {Pairam}},\ and\
  \bibinfo {author} {\bibfnamefont {A.}~\bibnamefont {Fernández-Nieves}},\
  }\href {https://doi.org/10.1088/1361-6463/ab08a7} {\bibfield  {journal}
  {\bibinfo  {journal} {Journal of Physics D: Applied Physics}\ }\textbf
  {\bibinfo {volume} {52}},\ \bibinfo {pages} {213001} (\bibinfo {year}
  {2019})}\BibitemShut {NoStop}%
\bibitem [{\citenamefont {Sussman}\ and\ \citenamefont
  {Beller}(2019)}]{dan_sussman}%
  \BibitemOpen
  \bibfield  {author} {\bibinfo {author} {\bibfnamefont {D.~M.}\ \bibnamefont
  {Sussman}}\ and\ \bibinfo {author} {\bibfnamefont {D.~A.}\ \bibnamefont
  {Beller}},\ }\bibfield  {journal} {\bibinfo  {journal} {Frontiers in
  Physics}\ }\textbf {\bibinfo {volume} {7}},\ \href
  {https://doi.org/10.3389/fphy.2019.00204} {10.3389/fphy.2019.00204} (\bibinfo
  {year} {2019})\BibitemShut {NoStop}%
\bibitem [{\citenamefont {Mottram}\ and\ \citenamefont
  {Newton}(2014)}]{Qtensor}%
  \BibitemOpen
  \bibfield  {author} {\bibinfo {author} {\bibfnamefont {N.~J.}\ \bibnamefont
  {Mottram}}\ and\ \bibinfo {author} {\bibfnamefont {C.~J.~P.}\ \bibnamefont
  {Newton}},\ }\href {https://doi.org/10.48550/ARXIV.1409.3542} {\bibinfo
  {title} {Introduction to q-tensor theory}} (\bibinfo {year}
  {2014})\BibitemShut {NoStop}%
\bibitem [{\citenamefont {Jackson}(1999)}]{Jackson}%
  \BibitemOpen
  \bibfield  {author} {\bibinfo {author} {\bibfnamefont {J.~D.}\ \bibnamefont
  {Jackson}},\ }\href@noop {} {\emph {\bibinfo {title} {Classical
  electrodynamics}}}\ (\bibinfo  {publisher} {Third edition. New York :
  Wiley},\ \bibinfo {year} {1999})\BibitemShut {NoStop}%
\bibitem [{\citenamefont {Ravnik}\ and\ \citenamefont {Zumer}(2009)}]{5cb}%
  \BibitemOpen
  \bibfield  {author} {\bibinfo {author} {\bibfnamefont {M.}~\bibnamefont
  {Ravnik}}\ and\ \bibinfo {author} {\bibfnamefont {S.}~\bibnamefont {Zumer}},\
  }\href@noop {} {\bibfield  {journal} {\bibinfo  {journal} {Liquid Crystals}\
  }\textbf {\bibinfo {volume} {36}},\ \bibinfo {pages} {1201} (\bibinfo {year}
  {2009})}\BibitemShut {NoStop}%
\end{thebibliography}%


\clearpage
\onecolumngrid
\renewcommand{\theequation}{S\arabic{equation}}
\setcounter{equation}{0}

\renewcommand{\thefigure}{S\arabic{figure}}
\setcounter{figure}{0}

\section*{Tunable Architecture of Nematic Disclination Lines \\ Supplementary Appendix}

\section*{Optical Setup}
\begin{figure*}[hb!]
\centering
\includegraphics[width= 0.5\linewidth]{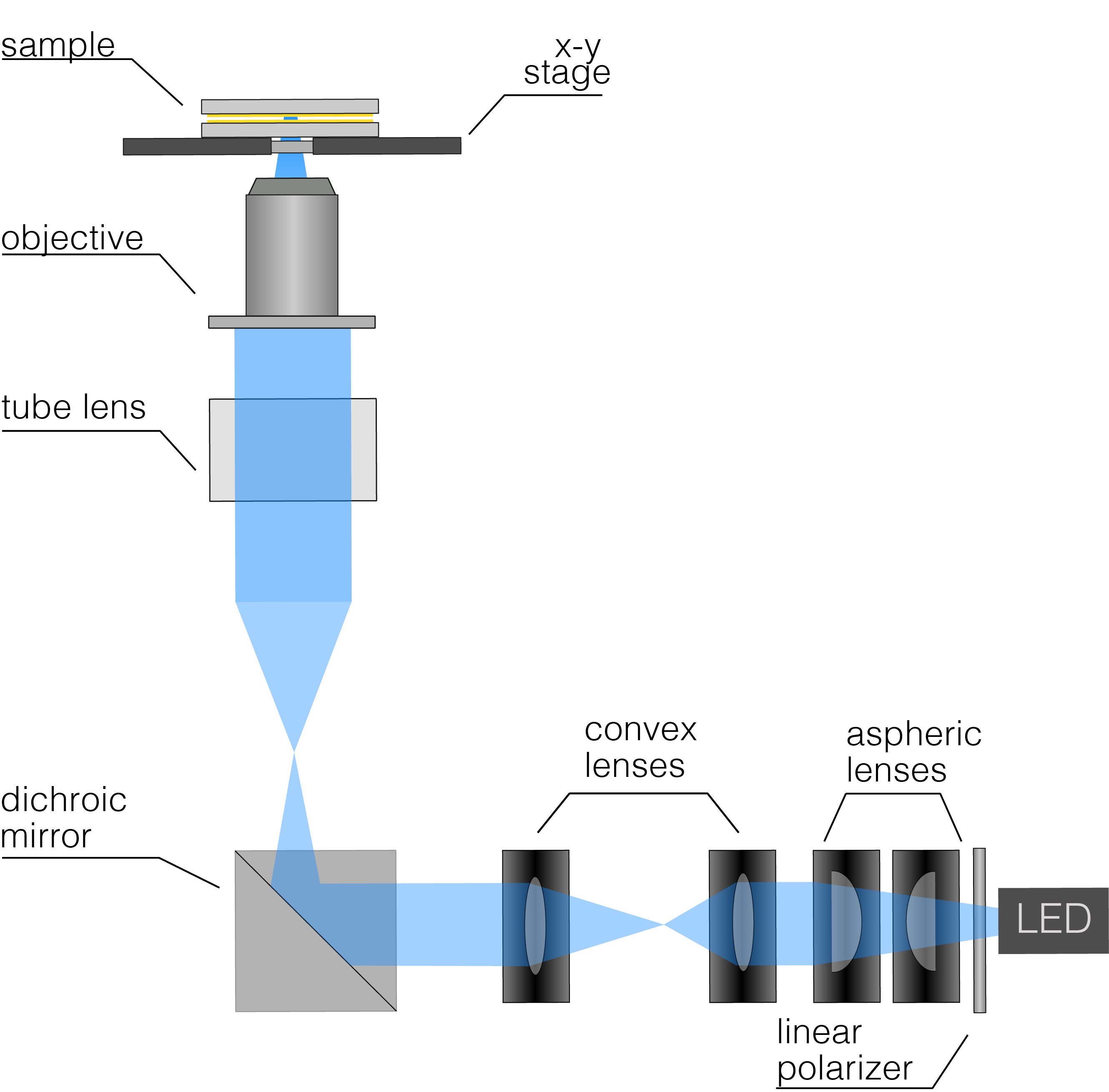}
\caption{\textbf{Optical setup for photoalignment.} A projector with blue LED light generates images focused on substrates at the sample plane. Images are polarized by a linear polarizer, collimated by aspheric lenses, and then expanded by a relay of lenses convex lenses. A microscope body contains a dichroic mirror that reflects light toward an infinity-connected tube lens. Images are picked up by the microscope and focused onto the sample plane.} 
\label{Fig:opticsDetails}
\end{figure*}

\section{Disclination lines connecting surface defects on opposing substrates}
\begin{figure*}[ht!]
\centering
\includegraphics[width= 0.9\linewidth]{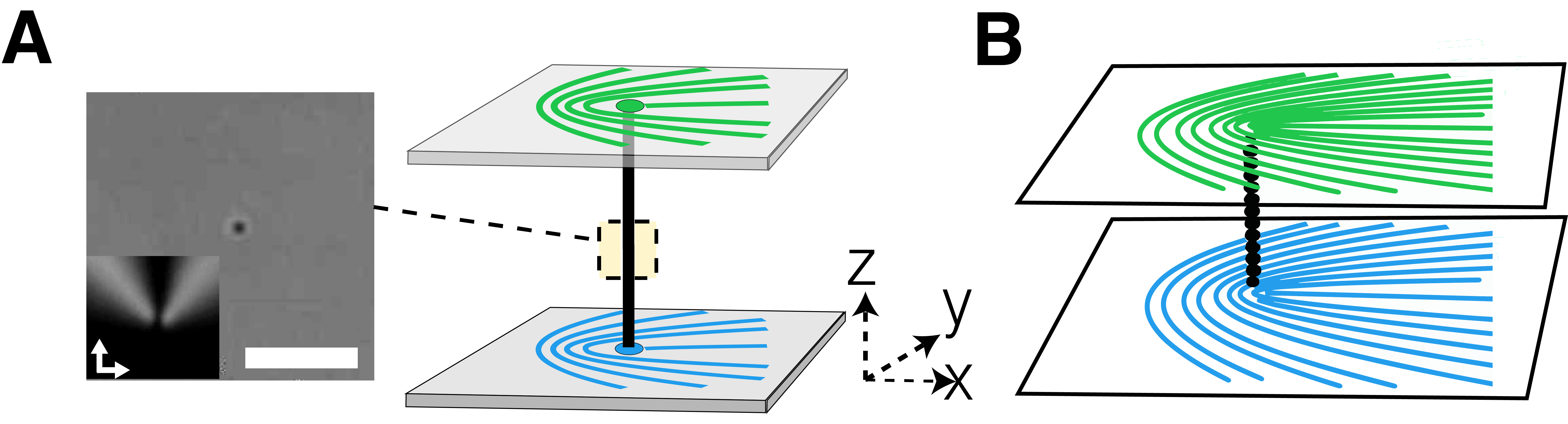}
\caption{\textbf{Defect connectivity with identical boundary conditions} (a) A single disclination line runs between surface defects with identical topological charges. In bright-field microscopy, the defect line appears as a single point at various locations within the LC cell. Inset: Polarized optical microscopy image. On the side, the schematic of the defect line. The image is taken in the mid-plane of the cell. Scale bar: 10 $\mu$m. (b) Simulation results of a straight disclination line between two surfaces with +1/2 point defects. }
\label{Fig:sameBoundary}
\end{figure*}

\section{Derivation of the forces acting on a wire element}
The magnetostatic model emerges from the similarity between equations [3,4] in the main text and their (vacuum) magnetostatics counterparts~\cite{Jackson}:
\begin{equation}\label{eq:Fmag}
	F_{mag}=\frac{1}{2\mu_0}\int\left|\mathbf{B}\right|^2 d\tilde{V}
\end{equation}
and
\begin{equation}\label{eq:disclination}
	\oint d\mathbf{\ell}\cdot \mathbf{B}= \mu_0 I
\end{equation}
By comparison, the following pairs are analogous:
\begin{table*}[h]
\centering
\begin{tabular}{|c|c|c|}
	\hline
	& Nematic & Magnetic \\
	\hline
	Field & $\nabla\theta$ & $\mathbf{B}$ \\
	\hline
	Modulus & $K$ & $1/\mu_0$ \\
	\hline
	Current & $2\pi K q$ & $I$ \\
	\hline
	Force per unit length on wire in field & $2\pi K q\,\mathbf{d\ell}\times\nabla\theta$ & $I\mathbf{d\ell}\times\mathbf{B}$ \\
	\hline
	Force per unit length between wires at distance $d$ & $2\pi K{q_1 q_2}/{d}$ & $\frac{\mu_0}{2\pi}{I_1 I_2}/{d}$ \\
	\hline
\end{tabular}
\end{table*}

The problem (namely, Eq.~\ref{eq:disclination} and the Euler-Lagrange equation associated with Eq.~\ref{eq:Fmag}) is linear. We may therefore write the boundary conditions as a sum of contributions, solve them separately, and add up the solutions/forces exerted in each case. 

We start with the method of images. We introduce an infinite ladder of mirror wires outside of our cell, indexed by $m\in\mathbb{Z}\setminus\{0\}$. We set the wires parallel to the actual one, located at $\tilde{z}_m=m\tilde{t}+(-1)^m\delta$, all carrying the same current $2\pi K q$. We now sum up the forces per unit length exerted on the wire element by its mirror images:
\begin{align}
	\mathbf{f}_M&=\sum_{\substack{m=-\infty \\ m\ne 0}}^{\infty}\frac{2\pi K q^2}{\delta-\tilde{z}_m}\mathbf{\hat{z}}=2\pi K q^2\sum_{\substack{m=1 \\ m~\textrm{odd}}}^{\infty}\left(\frac{\mathbf{\hat{z}}}{m\tilde{t}+2\delta}-\frac{\mathbf{\hat{z}}}{m\tilde{t}-2\delta}\right)\nonumber\\
	&=-\frac{\pi^2 K q^2}{\tilde{t}}\tan\left(\frac{\pi \delta}{\tilde{t}}\right)\mathbf{\hat{z}}.
\end{align}

By construction, the above (real and image) wire setup is symmetric about either of the two boundary plates. Therefore, $\nabla\theta$ (induced by this setup alone) is perpendicular to these boundaries, namely the director angles on both boundaries are constant (henceforth denoted $\theta_{t,b}^0$). By the integral condition in Eq.~\ref{eq:disclination} and by lateral reflection symmetry, $\theta_t^0-\theta_b^0=q\pi$. However, in the experimental/numerical setup discussed in the main text, $\Delta\theta=\theta_t-\theta_b$ is an arbitrary function of $x,y$.

To correct this, the solution to Eqs.~\ref{eq:Fmag} and \ref{eq:disclination} must be the sum of the above solution and a harmonic function (so that the equations are still satisfied) that makes up for the boundary condition mismatch. The (divergence-free) gradient of this function can be interpreted as an external magnetic field $\mathbf{B}_{ext}(x,y,z)$ that acts regardless of the exact shape of the wire. It can be written explicitly using Green's functions. However, we further assume that the thickness of the cell is much smaller than the lateral gradients of $\theta_{t,b}(x,y)$, therefore almost everywhere $\mathbf{B}_{ext}\approx B\mathbf{\hat{z}}$ (this assumption may fail close to surface defects). Being divergence-free, $B(x,y,z)$ must be approximately uniform in $z$, and matching the boundary conditions we get $B(x,y)=\frac{\Delta\theta-\Delta\theta^0}{\tilde{t}}=\frac{1}{\tilde{t}}(\theta_t-\theta_b-q\pi)$. Thus, the Lorenz force exerted on the wire by the external field reads
\begin{equation}
	\mathbf{f}_B=\frac{2\pi K q}{\tilde{t}} \left(\theta_{t}-\theta_{b}-q\pi\right)\mathbf{\hat{T}}\times\mathbf{\hat{z}},
\end{equation}
where $\mathbf{\hat{T}}$ is the unit tangent to the defect line.

Derivation of the force exerted by line tension is rather straightforward. A line segment of length $\delta$ is subject to tangential forces by its neighboring elements. The force per unit length is therefore,
\begin{equation}
	\mathbf{f}_\gamma=\lim\limits_{\delta\ell\to 0}\frac{\gamma\mathbf{\hat{T}}(\ell+\delta/2)-\gamma\mathbf{\hat{T}}(\ell-\delta/2)}{\delta}=\gamma\mathbf{\dot{\hat{T}}}=\gamma\kappa\mathbf{\hat{N}}
\end{equation}
where $\kappa,\mathbf{\hat{N}}$ are defined with the Frenet-Serret apparatus.

\section{Landau-de Gennes modeling of nematic liquid crystals}
In the Landau-de Gennes theory, the phenomenological free energy $F$ of a nematic liquid crystal can be written as~\cite{deGennesBook,dan_sussman}
\begin{align}
F=\int_{v} \left(f_{L} + f_{E}\right)dv + \int_{s} f_{B}ds,
\label{Eq.LG}
\end{align}
where $f_{L}$ is the Landau free energy density associated with deviation of the nematic order from its equilibrium value and can be expressed as~\cite{deGennesBook,dan_sussman}
\begin{align}
f_{L}=\frac{a}{2} \text{tr}(Q^2) + \frac{b}{3} \text{tr}(Q^3)  + \frac{c}{4} (\text{tr}(Q^2))^2.
\end{align}
Here, the phenomenological coefficients $a, b,$ and $c$ are nematic material parameters. To make the free-energy density dimensionless, all the energy terms are re-scaled by the energy scale $|a|$ in open-Qmin, which implies a non-dimensionalization of all the elastic constants. In the simulation, we take $a =-1.0, b =12.3,$ and $c=-10.0$~\cite{5cb}, which are commonly used in modeling of 5CB. Note that the values of $a, b,$ and $c$ determine the equilibrium mean-field value of the nematic order $S$ as~\cite{deGennesBook,dan_sussman}
\[
S = \frac{-b+ \sqrt{b^2-24ac}}{6c}.
\]
Under the two-constant approximation $K_1=K_3\equiv K \ne K_2$, and assuming strong anchoring at the boundaries (which render full-derivative terms irrelevant), the elastic free energy density $f_{E}$ is given by~\cite{dan_sussman}
\begin{equation}
f_{E}  = \frac{2K_2}{9s^2}\frac{\partial Q_{ij}}{x_k}\frac{\partial Q_{ij}}{x_k}
              +\frac{4K}{9s^2}\frac{\partial Q_{ij}}{x_j}\frac{\partial Q_{ik}}{x_k}
              -\frac{4K_2}{9s^2}\frac{\partial Q_{ik}}{x_j}\frac{\partial Q_{ij}}{x_k}
\end{equation}
For a broad temperature range within the nematic phase of $5CB$, the two-constant approximation is reasonable, and $K_2/K \approx 0.35$~\cite{elasticConsts}, which is the value we use in all our simulations. The last term $f_{B}$ in Eq.~\ref{Eq.LG} represents the energy density associated with the nematic directors at the boundary surface. We set the anchoring strength $W \gg |a|$ to achieve strong anchoring. We minimize the total free energy $F$ given in Eq.~\ref{Eq.LG} numerically with the values previously mentioned for different parameters using the lattice-discretized Landau-de Gennes modeling of nematic liquid crystals implemented in open-Qmin~\cite{dan_sussman}.

\section{Estimating $\frac{\gamma}{\sqrt{K K_2}}$ from the simulation} \label{sec.gammaK}
\begin{figure*}[t]
\centering
\includegraphics[scale=0.7]{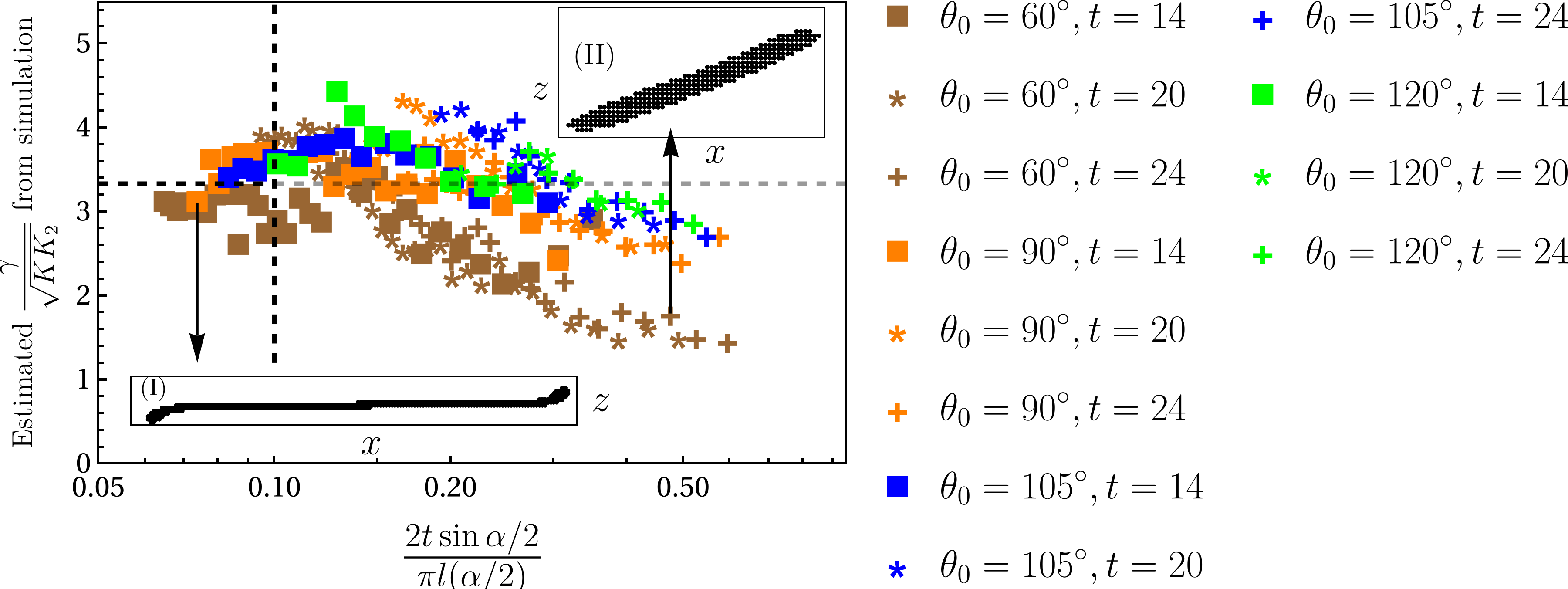}
\caption{Estimation of the parameter $\frac{\gamma}{\sqrt{K K_2}}$ in the simulation obtained by analyzing the circular arc-type defect configurations for a given $\theta_0$ and $t$ with different values of $l$. The $x$-axis is in logarithmic scale. The average value of the parameter $\frac{\gamma}{\sqrt{K K_2}}=3.3 \pm 0.1$ (horizontal dashed line in the plot) is obtained from the mean of all $\frac{\gamma}{\sqrt{K K_2}}$ for $\Tilde{x}=\frac{2}{\pi}\frac{t}{l} \frac{\sin\left(\frac{\alpha}{2} \right)}{\frac{\alpha}{2}}\le 0.1$. Insets show the defect configurations from the side view ($x$-$z$ plane where $z$-axis is along the thickness of the system) for two sets of parameters. Inset(I): Formation of a long horizontal line defect around the mid-plane for $\theta_0=90^{\circ}, l=145,$ and $ t=14$ which yields  $\alpha=2.7$, and $\tilde{\gamma} \sin\left(\frac{\alpha}{2}\right)=0.2$. Such a defect configuration satisfies the condition   $\tilde{\gamma} \sin\left(\frac{\alpha}{2} \right)\ll  \frac{\alpha}{2}$ and thus included in estimating the simulation parameter $\frac{\gamma}{\sqrt{K K_2}}$.
Inset(II): Side view of the defect configuration for $\theta_0=60^{\circ}, l=50,$ and $ t=24$ which yields  $\alpha=1.2$ and $\tilde{\gamma} \sin\left(\frac{\alpha}{2}\right)=0.4$. Instead of forming a long horizontal section around the mid-plane, the line defect changes continuously across the thickness of the system.}
\label{Fig:figGamma}
\end{figure*}

While we know $t, l, K,$ and $K_2$ in the simulation, $\gamma$ is not an input parameter. To estimate $\gamma$ in the simulation, we analyze the defect configurations for the boundary condition where two identical $+\frac{1}{2}$ defects are patterned on opposite surfaces (with a relative rotation of $\theta_0$ between the two patterns on the two surfaces) of a thin nematic cell of thickness $t$ with a separation $l$ between the defect centers (see main text for details). For a given $\theta_0$, we start the simulation with different random initial conditions for different values of $l$ and $t$ keeping $K$ and $K_2$ fixed. For a given $l, t, $ and $\theta_0$, when viewed from the top, the line defect in the energy minimized configuration forms a circular arc having opening angle $\alpha$. We determine $\alpha$ by fitting the circular arc with a circle under the constraint that the fitted circle must pass through the centers of the two defects. Note that $\sin\left(\frac{\alpha}{2} \right) = \frac{l}{2R}$, where $R$ is the radius of curvature of the fitted circle (see Fig.3B in the main text). Once we know $\alpha$ for a given $\theta_0$, we can estimate $\tilde{\gamma}$ from Eq.(9) mentioned in the main text. Using the definition of $\tilde{\gamma}=\frac{2}{\pi}\frac{\gamma t}{\sqrt{K K_2}l}$ in the same equation, we get
\begin{align}
\frac{\gamma}{\sqrt{K K_2}}&=\frac{\pi}{2}\frac{l}{t} \frac{\theta_0-\frac{\alpha}{2}}{\sin\left(\frac{\alpha}{2} \right)}.
\label{Eq.GamaK}
\end{align}
Thus, Eq.~(\ref{Eq.GamaK}) allows us to estimate $\frac{\gamma}{\sqrt{K K_2}}$ in the simulation for different $l, t,$ and $\theta_0$. In the simulation, we consider a system of size $250 \times 250 \times L_z$ with $L_z \in [15-25]$ and the separation between the defect centers $l \in [40-200]$. Note that thickness $t=L_z-1$. We find that $\frac{\gamma}{\sqrt{K K_2}}$ varies within a broad range for different $l, t,$ and $\theta_0$ as shown in Fig.~\ref{Fig:figGamma}. 

As discussed in the main text, Eq.~(\ref{Eq.GamaK}) is valid when the line defect forms a long horizontal section in the bulk, i.e., the lateral span $(\alpha R)$ of the line defect in the bulk should be much larger than the typical length scale $\frac{\gamma}{\sqrt{K K_2}}t$ set by the line tension. Thus, to estimate $\frac{\gamma}{\sqrt{K K_2}}$ from Eq.~(\ref{Eq.GamaK}) we need to consider only those defect configurations for which 
\begin{align}
\frac{\gamma}{\sqrt{K K_2}}t \ll \alpha R   \Rightarrow \frac{2}{\pi}\frac{t}{l} \frac{\gamma}{\sqrt{K K_2}} \sin\left(\frac{\alpha}{2} \right)=\tilde{\gamma} \sin\left(\frac{\alpha}{2} \right) \ll  \frac{\alpha}{2},
\end{align}
where, we have used $R= \frac{l}{2\sin\left(\frac{\alpha}{2} \right)}$. Thus, to estimate an optimal value of $\frac{\gamma}{\sqrt{K K_2}}$ in the simulation, we need to consider only those defect configurations for which $\tilde{x}=\frac{2}{\pi}\frac{t}{l} \frac{\sin\left(\frac{\alpha}{2} \right)}{\frac{\alpha}{2}} \ll 1$. Fig.~\ref{Fig:figGamma} shows the dependence of $\frac{\gamma}{\sqrt{K K_2}}$ on $\tilde{x}$ for different $t$ and $\theta_0$. To obtain an average value of $\frac{\gamma}{\sqrt{K K_2}}$, we consider the mean of all $\frac{\gamma}{\sqrt{K K_2}}$ for $\tilde{x} \le 0.1$ (vertical dashed line in the plot) which yields $\frac{\gamma}{\sqrt{K K_2}}=3.3 \pm 0.1$ (horizontal dashed line in the plot) for the simulation. As a self-consistency check, we compute $\tilde{\gamma} \sin\left(\frac{\alpha}{2} \right)=\frac{2}{\pi}\frac{t}{l} \frac{\gamma}{\sqrt{K K_2}} \sin\left(\frac{\alpha}{2} \right)$ for all the defect configurations considered in Fig.~\ref{Fig:figGamma} and find that defect configurations with $\tilde{x} \le 0.1$ satisfy the condition $\tilde{\gamma} \sin\left(\frac{\alpha}{2} \right)\ll  \frac{\alpha}{2}$ (for example see insets of Fig.~\ref{Fig:figGamma}). We find that a defect configuration with $\tilde{x} \le 0.1$ typically forms a long horizontal section around the mid-plane as shown in Fig.~\ref{Fig:figGamma}(I) for a particular set of parameters $(\theta_0=90^{\circ}, l=145,$ and $ t=14)$ while a defect configuration which does not form a horizontal section in the bulk typically yields $\tilde{x} > 0.1$. This observation allows us to estimate $\frac{\gamma}{\sqrt{K K_2}}$ in the simulation by considering all configurations with $\tilde{x} \le 0.1$.

\section{Comparison of simulation and experimental results} 
To compare the experimental and simulation results, we use the same value for $\tilde{\gamma}= \frac{2}{\pi}\frac{\gamma t}{\sqrt{K K_2}l}$. In the experiment, $t$ and $l$ are fixed, and we change temperature $(T)$, which changes the value of $\gamma$. In the simulation, we don't have $T$ directly. Thus, to mimic the role of $T$ (equivalently, $\gamma$) in the experiment, we can change $t$ and $l$ in the simulation in such a way that we have the same value for $\tilde{\gamma}$ in the experiment and simulation, i.e., we want
\begin{align}
\frac{2}{\pi} \frac{t_{e}}{l_{e}} \left(\frac{\gamma}{\sqrt{K K_2}}\right)_{e} = \frac{2}{\pi} \frac{t_s}{l_s} \left(\frac{\gamma }{\sqrt{K K_2}}\right)_{s},
\end{align}
where the subscript $s$ and $e$ represent the parameters for the simulation and experiment, respectively. Thus, to make a meaningful comparison, we need to have
\begin{align}
\left(\frac{l}{t}\right)_{s} = \left(\frac{l}{t}\right)_{e} \left(\frac{ \sqrt{K K_2}}{\gamma}\right)_{e}\left(\frac{\gamma }{\sqrt{K K_2}}\right)_{s}
\end{align}
In the experiment we have $l_e \approx 61 \mu m \pm 0.1 \mu m$ and $t_e=2.83 \mu m \pm 0.33 \mu m$. From the experimental results shown in Fig.~3(E), we find that for a broad temperature range $\left(\frac{\gamma}{\sqrt{K K_2}}\right)_{e} \in [12,18]$. As discussed in the previous section, we have that $\left(\frac{\gamma}{\sqrt{K K_2}}\right)_{s} = 3.3 \pm 0.1$ in the simulation. Thus, we find that in the simulation, we need to use $\left(\frac{l}{t}\right)_{s} \in [3.96, 5.94]$ which also implies $\tilde{\gamma} \in [0.35,0.53]$. 

For the heart-shaped pattern, we perform simulation for different $l$ and $t$ with $\tilde{\gamma} \in [0.35,0.53]$. In particular, we set $l=70$ lattice units and use $t=13$ and $15$ which gives $\tilde{\gamma}=0.39$ and $0.45$, respectively. To compare the obtained defect configuration with the experiment, we choose the temperature $T$ (see from Fig.3(E) in the main text) for which we have the same value of $\tilde{\gamma}$.

\end{document}